\documentclass[fleqn,usenatbib]{mnras}

\usepackage{newtxtext,newtxmath}

\usepackage[T1]{fontenc}

\usepackage{fix-cm}

\DeclareRobustCommand{\VAN}[3]{#2}
\let\VANthebibliography\thebibliography
\def\thebibliography{\DeclareRobustCommand{\VAN}[3]{##3}\VANthebibliography}

\usepackage{graphicx}	
\usepackage{amsmath}	
\usepackage{orcidlink}
\usepackage{comment}
\usepackage{url}

\defcitealias{yangGalaxyGroupsSDSS2007}{Y07}
\defcitealias{hsuSDSSIVMaNGACannibalism2022}{Hsu22}

\title[Density-Based Galaxy Group Finder]
{sOPTICS: A Modified Density-Based Algorithm for Identifying Galaxy Groups/Clusters and Brightest Cluster Galaxies}

\author[H.-X.\ Ma et al.]{
Hai-Xia Ma$^{1}$\thanks{E-mail: \href{mailto:ma.haixia.k8@s.mail.nagoya-u.ac.jp}{ma.haixia.k8@s.mail.nagoya-u.ac.jp}} \orcidlink{0000-0002-5237-9433},
Tsutomu~T.~Takeuchi$^{1,2}$\orcidlink{0000-0001-8416-7673}, 
Suchetha~Cooray$^{3, 4}$\orcidlink{0000-0002-9217-1696},
Yongda Zhu$^{5}$\orcidlink{0000-0003-3307-7525}
\\
$^{1}$ Division of Particle and Astrophysical Science, Nagoya University, Furo-cho, Chikusa-ku, Nagoya 464–8602, Japan\\
$^{2}$ The Research Center for Statistical Machine
Learning, the Institute of Statistical Mathematics, 10--3 Midori-cho, Tachikawa, Tokyo 190--8562, Japan\\
$^{3}$ Kavli Institute for Particle Astrophysics and Cosmology, Stanford University, 452 Lomita Mall, Stanford, CA 94305, USA\\
$^{4}$ SLAC National Accelerator Laboratory, 2575 Sand Hill Road, Menlo Park, CA 94025, USA\\
$^{5}$ Steward Observatory, University of Arizona, 933 N Cherry Ave, Tucson, AZ 85721, USA}

\date{Accepted XXX. Received YYY; in original form ZZZ}

\pubyear{2025}

\begin{document}
\label{firstpage}
\pagerange{\pageref{firstpage}--\pageref{lastpage}}
\maketitle

\begin{abstract}
A direct approach to studying the galaxy-halo connection is to analyze groups and clusters of galaxies that trace the underlying dark matter halos, emphasizing the importance of identifying galaxy clusters and their associated brightest cluster galaxies (BCGs). 
In this work, we test and propose a robust density-based clustering algorithm that outperforms the traditional Friends-of-Friends (FoF) algorithm in the currently available galaxy group/cluster catalogs. 
Our new approach is a modified version of the Ordering Points To Identify the Clustering Structure (OPTICS) algorithm, which accounts for line-of-sight positional uncertainties due to redshift space distortions by incorporating a scaling factor, and is thereby referred to as sOPTICS. 
When tested on both a galaxy group catalog based on semi-analytic galaxy formation simulations and observational data, our algorithm demonstrated robustness to outliers and relative insensitivity to hyperparameter choices.
In total, we compared the results of eight clustering algorithms. The proposed density-based clustering method, sOPTICS, outperforms FoF in accurately identifying giant galaxy clusters and their associated BCGs in various environments with higher purity and recovery rate, also successfully recovering 115 BCGs out of 118 reliable BCGs from a large galaxy sample. Furthermore, when applied to an independent observational catalog without extensive re-tuning, sOPTICS maintains high recovery efficiency, confirming its flexibility and effectiveness for large-scale astronomical surveys.
\end{abstract}




\section{Introduction}
\label{sec:intro}

Galaxy groups are fundamental structures in the universe comprising multiple galaxies bound together by gravity within a dark matter halo \citep{whiteCoreCondensationHeavy1978}. Galaxies in a group are located near the peak of this dark matter density distribution, where the gravitational potential is deepest \citep{mooreResolvingStructureCold1998, thanjavurDARKMATTERDISTRIBUTION2010, shinSpatialDistributionDark2022}. More numerous aggregations of galaxies are classified as clusters of galaxies, composed of hundreds to thousands of galaxies, hot gas, and predominantly dark matter. Galaxy groups and clusters are key components in understanding the formation of hierarchical structures in the universe, especially since they are closely related to dark matter halos. 

Therefore, identifying groups and clusters of galaxies is a crucial step in understanding the distribution and evolution of matter in the universe. The study of galaxy groups and clusters has been an active field of research for decades, with various methods developed for identifying and characterizing these structures.

In addition, central to galaxy clusters are the Brightest Cluster Galaxies (BCGs) located at the bottom of gravitational well within the clusters \citep{quintanaDeterminationVelocityDispersions1982}. The properties of BCGs dictate cluster formation and evolution, where BCG mass growth is closely tied to the hierarchical assembly and dynamical state of the host galaxy cluster \citep{sohnHectoMAPClusterSurvey2021}. What is more distinct from other galaxies is that some of the BCGs show multiple nuclei \citep[e.g.][]{lauerMorphologyMultipleNucleusBrightest1988, klugeStructureBrightestCluster2020}, making them good systems to study about galactic mergers. A recent study on velocity dispersion profiles of elliptical galaxies also found the majority of the BCGs exhibit flat velocity dispersion profiles \citep{tianMassVelocityDispersion2021, duannClassifyingMaNGAVelocity2023}. A distinct radial acceleration relation (RAR) has even been identified in BCGs, making them essential to study as they pose a significant challenge to the cold dark matter (CDM) paradigm \citep{tianDistinctRadialAcceleration2024}. However, identifying BCGs can be complex, often requiring a comprehensive survey of galaxies and coherently identifying a pure and complete galaxy clusters catalog first.

To effectively identify galaxy groups, we need to identify denser regions within a sparse distribution of galaxies. This approach resembles finding concentrated islands amid a vast, sparse ocean. 
Traditionally, the foundation of clustering algorithms has been based on the single-link clustering methodology. A quintessential example of this approach is the Friends-of-Friends \citep[FoF; ][]{turnerGroupsGalaxiesCatalog1976, huchraGroupsGalaxiesNearby1982, pressHowIdentifyWeigh1982, tagoGroupsGalaxiesSDSS2008} algorithm. The FoF method links galaxies within a specified proximity, progressively forming clusters. Single-link clustering generally results in clusters where even distantly related members are interconnected through a sequence of nearer members, leading to the term "single-link". However, this method is sensitive to noise, where an isolated noisy data point might erroneously connect neighbor clusters, leading to the merging of clusters that are otherwise distinct, compromising the accuracy of the clustering results. Additionally, this approach can yield clusters with a chain-like configuration, which is highly sensitive to the predefined linking length, a hyperparameter. Nevertheless, despite these limitations, single-link clustering remains a valuable tool due to its efficiency and simplicity, particularly for identifying groups and clusters of stars or galaxies with elongated or irregular shapes \citep{sankhyayanIdentificationSuperclustersTheir2023, chiBlindSearchSolar2023}.

Density-based clustering methodologies, such as the Density-Based Spatial Clustering of Applications with Noise \citep[DBSCAN; ][]{esterDensitybasedAlgorithmDiscovering1996, sanderDensityBasedClusteringSpatial1998}, have been introduced to address these limitations and enhance the robustness of cluster identification. These methods rely on estimating the density of data points, allowing for separation between lower-density areas and higher-density regions. The primary aim here is not to distinctly separate these two areas but to enhance the robustness of the identified core clusters against noise. By doing so, these algorithms provide a more reliable means of cluster identification, which is crucial in analyzing galaxy distributions. Therefore, density-based clustering algorithms for identifying galaxy groups have emerged as alternatives to FoF. DBSCAN identifies clusters based on the density of points, designating core points with a high density of neighbors and expanding clusters from these cores. This method effectively lowers the influence of isolated noise points, thus making the identification of clusters of points more robust and reflective of the true spatial distribution. Its effectiveness is particularly notable for discovering open clusters of stars \citep{castro-ginardNewMethodUnveiling2018, castro-ginardHuntingOpenClusters2020} as well as clusters and groups of galaxies \citep{dehghanClustersGroupsFilaments2014, olave-rojasCALSAGOSClusteringAlgorithms2023}.
 
However, DBSCAN has limitations, particularly in handling datasets with clusters of varying densities. Since it relies on a single density threshold to define clusters, DBSCAN can struggle to effectively identify clusters with differing density levels. To address these shortcomings, algorithms such as Hierarchical Density-Based Spatial Clustering of Applications with Noise \citep[HDBSCAN; ][]{campelloHierarchicalDensityEstimates2015, mcinnesHdbscanHierarchicalDensity2017} and Ordering Points To Identify the Clustering Structure \citep[OPTICS; ][]{ankerstOPTICSOrderingPoints1999} have been introduced. These algorithms improve upon DBSCAN by adapting to local density variations. In both HDBSCAN and OPTICS, the process can be visualized as ‘lowering the sea level’ in a topographical representation of the dataset: sparse points, which correspond to noise, are gradually pushed away from denser regions (the ‘land’), spreading them out as the sea level drops. Meanwhile, the denser regions, representing clusters, remain largely unchanged, as they stand above the receding sea level. This approach ensures that noise is separated from clusters, allowing for the identification of groups of stars and galaxies that better reflect the real distribution and density variations.
\citep{brauerPossibilitiesLimitationsKinematically2022, fuentesStellarHaloHierarchical2017, oliverHierarchicalStructureGalactic2021}.

Since BCGs are typically located at the bottom of the gravitational well, often indicating the densest region of a galaxy cluster, density-based clustering methods are anticipated to be particularly effective for identifying BCGs, even in complex and noisy environments. This effectiveness arises from the inherent capability of these methods to concentrate on the most dense regions, provided that the corresponding hyperparameters are set appropriately to define BCGs. In contrast, the FoF algorithm may struggle with clustering galaxies upon varying density. This is because its criteria for linking galaxies do not rely on density but on proximity, which might not accurately reflect the underlying density variations, especially in identifying BCGs.

Therefore, in this work, we conduct comprehensive tests on various clustering methods to explore the possibilities and challenges of identifying galaxy groups and clusters from large galaxy surveys and propose a new algorithm. 
This paper is organized as follows: 
Section~\ref{sec:method} provides a concise introduction to the clustering algorithms used in this study, detailing the methodology for feature extraction and hyperparameter optimization, including the selection criteria. 
Section~\ref{sec:test-simulation} offers a comprehensive evaluation of the effectiveness of group finders using a galaxy catalog derived from simulations. Following this, Section~\ref{sec:test-SDSS} presents additional tests conducted with real-world observations, which include mitigating redshift space distortion using our proposed line-of-sight scaling factor and comparisons with a reliable group catalog. Section~\ref{sec:discussion} discusses the strength of our sOPTICS method and its efficiency in identifying galaxy groups/clusters and BCGs. Finally, Section~\ref{sec:summary} summarizes our findings and provides a detailed discussion of the results.

\section{Clustering Methodology}
\label{sec:method}

Clustering, a key machine learning technique, groups similar data points into clusters and is a powerful tool for identifying galaxy groups/clusters in astronomical data. 
Among the available clustering algorithms, the most straightforward and widely used is $k$-means \citep{macqueenMethodsClassificationAnalysis1967}, often favored for their simplicity and computational efficiency in partitioning data into distinct groups. Beyond this foundational method, more advanced algorithms, such as the Gaussian Mixture Models (GMMs) \citep{dempsterMaximumLikelihoodIncomplete1977}, Spectral Clustering \citep{ngSpectralClusteringAnalysis2001, vonluxburgTutorialSpectralClustering2007} and Agglomerative Clustering \citep{wardHierarchicalGroupingOptimize1963}, are frequently employed across diverse domains due to their flexibility in identifying more complex or hierarchical data structures. They all share the critical hyperparameter of the number of clusters, $ N_\mathrm C$.

However, notably, a more advanced algorithm may not always be the optimal choice for every dataset or task, as it heavily depends on the specific use case and data type. Therefore, algorithm selection should prioritize suitability for the problem, especially when extracting galaxy groups/clusters from observations.
Thus, a comprehensive comparative study testing the suitability of various clustering algorithms is essential to identify the most effective approaches for this task.

\subsection{Fiducial Clustering Algorithms}
\label{sec:algorithms}

Before presenting the results of applying these algorithms to identify galaxy groups and clusters, we briefly describe the fiducial algorithms employed in our analysis.

\subsubsection{Friends-of-Friends (FoF)}

In the specific context of identifying galaxy groups or clusters based on their spatial distributions, the Friends-of-Friends (FoF) algorithm stands out as the most commonly used approach. Its popularity arises from its simplicity and its inherent ability to effectively capture the hierarchical and complex characteristics of cosmic structures. Given a set of points in space, the FoF algorithm links points within a predetermined distance $l$ to identify interconnected clusters. Two points are considered ‘friends’ (i.e., part of the same cluster) if they are within $l$ of each other. This process iteratively groups points together by linking points to their friends and friends of friends.

The FoF method critically relies on the linking length, $l$, to define the scale at which structures are identified. That is, smaller values of $l$ lead to the identification of smaller, tighter clusters by focusing on locally dense regions, but may fail to connect broader structures. Conversely, larger values of $l$ allow for the detection of more extended, diffuse clusters by linking points over greater distances. However, overly large $l$ values risk merging distinct structures into a single cluster, reducing the purity of the results. As a result, careful tuning of $l$ is crucial to balance the trade-off between identifying small-scale structures and preserving the integrity of larger-scale groupings.

\subsubsection{DBSCAN}

As mentioned in Section \ref{sec:intro}, density-based clustering algorithms such as DBSCAN, HDBSCAN, and OPTICS have been relatively underutilized or insufficiently explored in the context of identifying galaxy groups or clusters. However, these methods hold significant potential due to their capability to identify clusters of arbitrary shapes and effectively handle noise in the data. In this work, we focus on an in-depth investigation of these algorithms, examining their suitability and performance for extracting galaxy groups and clusters from observational datasets.

For each point in the dataset, DBSCAN calculates the number of points within a specified radius $\epsilon$. If this number exceeds a minimum number of neighbors $N_\mathrm{min}$, the point is classified as a core point, indicating a high-density area surrounding it. 
The core distance, $d_\mathrm{core}(P, N_\mathrm{min})$, is the distance from a point P to its $N_\mathrm{min}$-th most nearest neighbor and if $d_\mathrm{core}(P, N_\mathrm{min}) \leq \epsilon$ then P is a core-point.
These core points serve as the seeds for cluster growth, as the algorithm iteratively adds directly reachable points (points located within the $\epsilon$-radius of a core point) to their respective clusters. Points not reachable from any core point are labeled as noise.

For DBSCAN, the two primary hyperparameters are $N_\mathrm{min}$ and $\epsilon$, which together determine how a point qualifies as a core point. Larger $\epsilon$ values allow for the inclusion of more distant galaxies within clusters, enabling the identification of larger, more extended groups but increasing the risk of over-grouping due to projection effects. Conversely, higher $N_\mathrm{min}$ thresholds help identify more substantial clusters by focusing on denser regions, reducing the likelihood of detecting spurious or minor groupings. Specifically in this work, we introduce an additional layer of complexity with the \texttt{min\_member} parameter ($M_\mathrm{min}$) to DBSCAN. This parameter specifies the minimum number of members required for a grouping to qualify as a valid cluster, enabling the algorithm to filter out insignificant or noisy structures effectively. The $M_\mathrm{min}$ parameter is also applied in HDBSCAN and OPTICS in the subsequent discussion.

\subsubsection{HDBSCAN}

HDBSCAN, on the other hand, builds upon DBSCAN's concept but introduces a hierarchy of clusters. It first estimates the density of each point using the mutual reachability distance, 
\begin{equation}
    d_\mathrm{reach}(P, Q) = \max\{d_\mathrm{core}(P, N_\mathrm{min}), d_\mathrm{core}(Q, N_\mathrm{min}), d(P, Q)\}\,,
\end{equation}
where $d(P, Q))$ is the Euclidean distance between two points. HDBSCAN then constructs a minimum spanning tree (MST: e.g., \citealt{fouldsGraphTheoryApplications1992}), which connects all data points in a way that the total sum of edge lengths (distances) is minimized. By systematically removing the longest edges from the MST, HDBSCAN creates a dendrogram that reflects the data structure at varying density levels. Each cluster's stability is calculated as the sum of the excess of density (over a minimum cluster size threshold) for each point within the cluster across the range of distance scales. Finally, HDBSCAN iteratively prunes this dendrogram using the stability criterion, resulting in robust and persistent clusters over a range of densities.
\citep{campelloDensityBasedClusteringBased2013}

Comparing to DBSCAN, HDBSCAN incorporates another key hyperparameter $\alpha$, which governs the minimum stability a cluster must achieve to be considered significant. Lower values of $\alpha$ reduce the stability threshold, making it easier for points to be included in a cluster. While this can lead to larger and less dense clusters, it may also increase the likelihood of capturing subtle structures at the cost of potential over-grouping.

\subsubsection{OPTICS}

Instead of relying solely on a global $\epsilon$ parameter, OPTICS uses reachability to create an ordered list that reflects the data structure. The reachability distance in OPTIC is defined as: 
\begin{equation}
    d_\mathrm{reach}(P, Q) = \max\{d_\mathrm{core}(P, N_\mathrm{min}), d(P, Q)\}\,,
\end{equation}
which differs slightly from HDBSCAN’s approach, as it is specifically designed to construct the ordered list.
The ordered list is built by iteratively calculating and updating the reachability distance for each data point. Starting with an arbitrary point, its reachability distance is calculated relative to its neighbors. This point is added to the list, and the algorithm progresses to the unprocessed point with the smallest reachability distance. This process continues until all points are ordered. The resulting list encapsulates the density-based clustering structure without explicitly assigning points to clusters.
Clusters can be then extracted by identifying valleys (low reachability distances, representing dense areas) separated by peaks (high reachability distances, marking transitions between clusters or noise) in the reachability plot. The $\xi$ parameter defines what constitutes a “steep” change in the reachability plot: it is used to detect sharp transitions (e.g., steep downward or upward slopes) by comparing the relative change in reachability distance between consecutive points in the ordered list. Specifically, $\xi$ marks the boundaries of these changes, identifying where clusters start or end, rather than marking all points within a region of steep gradient. This approach ensures that the clustering structure is captured at multiple density scales while avoiding over-segmentation of data.

Similar to DBSCAN, the parameter $N_\mathrm{min}$, which defines the minimum number of neighbors required to consider a point as a core point, controls the algorithm’s sensitivity to local density variations. Smaller values of $N_\mathrm{min}$ allow for the identification of smaller, more localized clusters, but may also increase the detection of noise and spurious groupings. The $\epsilon$ parameter determines the maximum distance for evaluating reachability, and while it does not directly dictate cluster boundaries as in DBSCAN, it sets an upper limit for defining local neighborhoods. Finally, the $\xi$ parameter, which identifies steep changes in the reachability plot, governs the resolution of cluster extraction. Lower $\xi$ values detect finer density variations, potentially identifying smaller or closely spaced clusters, whereas higher $\xi$ values focus on broader, more prominent clustering structures. Together, these parameters provide OPTICS with the flexibility to adapt to various density scales and cluster complexities.

\subsection{Hyperparameter Optimization}
\label{sec:hyperparameter}

All clustering algorithms in this work require a preselected hyperparameter carefully considered for the desired galaxy group scale. Choosing the optimal hyperparameter values involves balancing the preservation of large-scale structures against the fragmentation of real galaxy groups into smaller, potentially insignificant groups. To optimize hyperparameter values, we adopt two classical criteria, purity and completeness, to evaluate the performance of clustering algorithms under different hyperparameter settings. 
However, comparing results to a simulated group catalog introduces inherent biases. Simulations, while valuable, are not perfect representations of real galaxy groups, and even semi-analytic galaxy group catalogs are usually constructed using FoF clustering algorithms \citep[][also see Section \ref{sec:data-simulation}]{onionsSubhaloesGoingNotts2012}, introducing bias in the "ground truth" data. Consequently, demanding complete overlap between predicted and simulated groups is unrealistic and unnecessary. Instead, similar to \citet{brauerPossibilitiesLimitationsKinematically2022}, 
we define a broader measure of purity and completeness,  incorporating what we term as \textit{soft criteria}, to assess the performance of the clustering algorithms for a more nuanced evaluation.
Under these criteria, a cluster is considered pure if at least two-thirds of its galaxies originated from a single group and complete if it contains at least half the galaxies from that originating group.
Building upon the definitions of purity and completeness, we can define the purity rate and recovery rate for all predicted groups relative to the full set of true groups:
\begin{equation}
    F_{\mathrm{P}} = 
    \frac{\text{Number of pure groups}}
    {\text{Total number of predicted groups}}\,,
\end{equation}
\begin{equation}
    F_{\mathrm{R}} = 
    \frac{\text{Number of simultaneously pure and complete groups}}
    {\text{Total number of true groups}}\;.
\end{equation}
When calculating purity and recovery rates, we only compare predicted groups to true groups exceeding the minimum member threshold, $M_\mathrm{min}$=5. It is worth noting that, to calculate purity, 
the traversal list is defined as all the predicted groups in the clustering results, not the true groups in the simulation catalog. Specifically, for calculating the purity rate, we evaluate each predicted group to determine whether its members correspond to a single true group in the simulation catalog, rather than starting from the simulation catalog to find the corresponding predicted group. Conversely, for calculating the recovery rate, the traversal list consists of all halos with at least $M_\mathrm{min}$ galaxy members in the simulation catalog, rather than the predicted groups.
As a result, the purity rate reflects the proportion of predicted groups whose members are exclusively from a single true group, providing confidence that the algorithm correctly groups members together. A high purity rate indicates the algorithm’s effectiveness in identifying true groupings.
Consequently, the purity rate reflects the proportion of predicted groups exclusively containing members from a single true group. It provides confidence for ensuring that the members within a predicted group are genuinely bound together. A high purity rate indicates that the algorithm is effective in correctly grouping members. On the other hand, the recovery rate measures the percentage of true, significant groups successfully identified and reproduced by the algorithm. This ensures the informativeness and reliability of the results for further analysis.

We define search spaces of approximately 20 trial values for each hyperparameter to explore the impact of various hyperparameter choices. We then execute the clustering algorithm with each set of trial values and calculate purity and recovery rates (see Section \ref{sec:test-simulation}). The optimal hyperparameter values for each algorithm are chosen by maximizing the recovery rate. In cases where multiple sets yield the same recovery rate, the set with the highest purity rate is preferred. Table \ref{tab:hypervalues} provides an overview of the trial hyperparameter values and the optimized results against the simulated group catalog.

\section{Tests with simulated group catalog}
\label{sec:test-simulation}

A crucial test of any group finder's performance involves comparing its results to the expected distribution of galaxies in a group catalog built from simulations using semi-analytic models \citep[SAMs, ][]{kauffmannUnifiedModelEvolution2000, springelPopulatingClusterGalaxies2001}. This allows us to assess how well the group finder aligns with the theoretical framework of galaxy formation. In this work, we utilize a galaxy group catalog \citep{crotonManyLivesActive2006, deluciaHierarchicalFormationBrightest2007} built from The Millennium Simulation \citep{springelSimulationsFormationEvolution2005}, which provides a well-established and widely used benchmark for testing group finder performance.

\subsection{Galaxy Sample}
\label{sec:data-simulation}

The Millennium Simulation tracks the evolution of $N= 2160^3$ dark matter particles within a comoving volume of $500\, \mathrm{h}^{-1}\mathrm{Mpc}$ using the N-body code \texttt{GADGET-2} \citep{springelCosmologicalSimulationCode2005}. Sixty-four snapshots were periodically saved, along with group catalogs and their substructures identified through a two-step process. First, the FoF algorithm with a linking length of 0.2 in units of the mean particle separation identified potential halos. These candidates were then refined by the \texttt{SUBFIND} algorithm \citep{springelPopulatingClusterGalaxies2001} through a gravitational unbinding procedure, ensuring only substructures with at least 20 particles were considered to be genuine halos and substructures. Subsequently, with halos detailed merger history trees were constructed for all gravitationally bound structures in each snapshot. The merger trees trace the evolution of these structures throughout cosmic time, providing the crucial temporal and structural framework upon which SAMs operate. Within this framework, SAMs simulate the formation and evolution of galaxies, ultimately populating the dark matter halos with galaxies \citep{deluciaHierarchicalFormationBrightest2007}.

From the semi-analytic galaxy group catalogs of \citet{deluciaHierarchicalFormationBrightest2007}, we extracted a cubic sub-volume of $50\, \mathrm{h}^{-1}\mathrm{Mpc}$ side length at snapshot 63 (corresponding to redshift $z=0$) as our fiducial test sample. This sub-volume contains 26,276 galaxies originating from 17,878 halos. However, most of these halos only host a single galaxy, making them unsuitable for characterizing groups or clusters. Therefore, we focused on halos containing at least five galaxies, resulting in a final sample of 400 halos. We further expanded our analysis by extracting similar sub-volumes at snapshots 30 ($z=2.422$), 40 ($z=1.078$), and 50 ($z=0.408$) to explore the performance of the clustering algorithms across different cosmic epochs. It is important to note that our analysis is restricted to real space (3D Cartesian coordinates) for computational efficiency, neglecting the effects of peculiar velocities and, consequently, redshift-space distortions.

\subsection{Comparing Clustering Algorithms}
\label{sec:result-simulation}

Considering the hierarchical structure of the Universe, with galaxy groups typically hosting 3 to 30 bright galaxies and clusters holding 30 to over 300, it is 
logical to focus our search for optimal clustering parameters within this range. The Local Group, for instance, hosts over 30 galaxies with a diameter of nearly 3 Mpc \citep{mcconnachieDistancesMetallicities172005}. 
Therefore, we set the upper limit for the linking length $l$ in FoF, $\epsilon$ in OPTICS and DBSCAN to 3.0 Mpc, corresponding to a reasonable maximum for the spatial sparsity of galaxies within a group.
Similarly, the minimum neighbor number $N_\mathrm{min}$ and minimum member number $M_\mathrm{min}$ are explored within the range of 2-20. We employ a broader range, 500 to 5000, for algorithms requiring a preselected number of clusters, to ensure exploring all possibilities. The trial hyperparameter values for all algorithms are listed in  Table \ref{tab:hypervalues}.

We apply the eight algorithms described in Section \ref{sec:method} to the test sample obtained in Section \ref{sec:data-simulation}, evaluating each algorithm with all trial hyperparameter values. The \texttt{python} package \texttt{GalCluster} we developed to conduct the tests is realized. This tool lets users easily perform galaxy group finding on a simulated observed catalog. We calculate the purity and recovery rates for each run according to the soft criteria by comparing the predicted groups with the true halo IDs in the simulation. We subsequently select the optimal hyperparameters that maximize the recovery rate. The complete results, including the predicted groups corresponding to the optimal hyperparameters, are presented in Table \ref{tab:hypervalues}.

{
\renewcommand{\arraystretch}{1.2}
\begin{table*}
\begin{tabular}{ccccccc}
\hline
Algorithm  & Hyperparameter & Search space & Optimal Value & Maximum recovery rate  & Number of groups\\         \hline
Friend-of-friends & \texttt{linking\_length} $l$ &  [0.001, 3.0) & 0.24 & 79.0\% & 471 \\ \hline
        & \texttt{max\_eps} $\epsilon$       & [0.05, 3.0) & 0.25 & & \\
OPTICS  & \texttt{min\_sample} $N_\mathrm{min}$ & [2, 20)     & 3  &  79.2\% & 472\\
        & \texttt{xi} $\xi$         & [0.05, 1.0)   & 0.95 & & \\
        & \texttt{min\_member} $M_\mathrm{min}$ & [5, 20) & 5 & & \\
        \hline
DBSCAN  & \texttt{max\_eps} $\epsilon$ & [0.05, 3.0)     & 0.2  & & \\
        & \texttt{min\_sample} $N_\mathrm{min}$ & [2, 20)     & 2 &  75.2\% & 406\\
        & \texttt{min\_member} $M_\mathrm{min}$ & [5, 20) & 5 & & \\
\hline
        & \texttt{min\_sample} $N_\mathrm{min}$  & [2, 20) & 2 & & \\
HDBSCAN & \texttt{min\_member} $M_\mathrm{min}$  & [5, 20) & 5 & 49.5\% & 1576\\
        & \texttt{alpha} $\alpha$ & [0.05, 1]   & 0.90 & &\\
        \hline
$k$-means &    &  & >5000 & 54.7\% & 5000 \\
GMMs  & \texttt{n\_clusters} $N_\mathrm{C}$ & [500, 5000) & >5000 & 7.6\%& 5000\\
Agglomerative Clustering &   &  & >5000 & 57.3\% & 5000\\ 
Spectral Clustering &  &  & too slow & - & - \\
\hline
\end{tabular}
\caption{Trial hyperparameter values for all algorithms and the best fitting values to maximize the recovery rate under soft criteria, respectively. Note that since the processing time for GMM, $k$-means, Agglomerative Clustering, and Spectral Clustering on the test sample are very long, and Iterating them over the searching space takes even more time, we hereby adopt a subsample for fitting their hyperparameters.}
\label{tab:hypervalues}
\end{table*}
}

As we can see from the results, the traditional methods FoF, OPTICS, and DBSCAN can effectively recover the galaxy groups just based on the spatial distribution of galaxies with a recovery rate of over 70\%. The others, including HDBSCAN, can not give a good prediction of the groups. It should also be emphasized that the ground truth was calculated based on the FoF algorithm. 

In particular, $k$-means, GMMs, Agglomerative Clustering, and Spectral Clustering all demonstrate significant shortcomings in this context. Their results clearly show that they are not suitable for identifying galaxy clusters. This is due to the nature of the algorithms themselves: they assume fixed or simplistic cluster shapes (e.g., spherical or Gaussian) and rely on predefined scales, which do not align with the complex, hierarchical, and irregular distribution of galaxies in cosmic structures. What is worse is that Spectral Clustering not only struggles with the complexity of the task but is also computationally expensive. Even for a moderately sized sample much smaller than real observational datasets, the algorithm required excessive runtime to complete, rendering it impractical for larger-scale applications.
Consequently, we conclude that these four algorithms are unsuitable for finding galaxy clusters and will not be included in the following tests or discussions.

\subsection{Parameter Sensitivity}

\begin{figure*}
    \centering
    \includegraphics[width=1\linewidth]{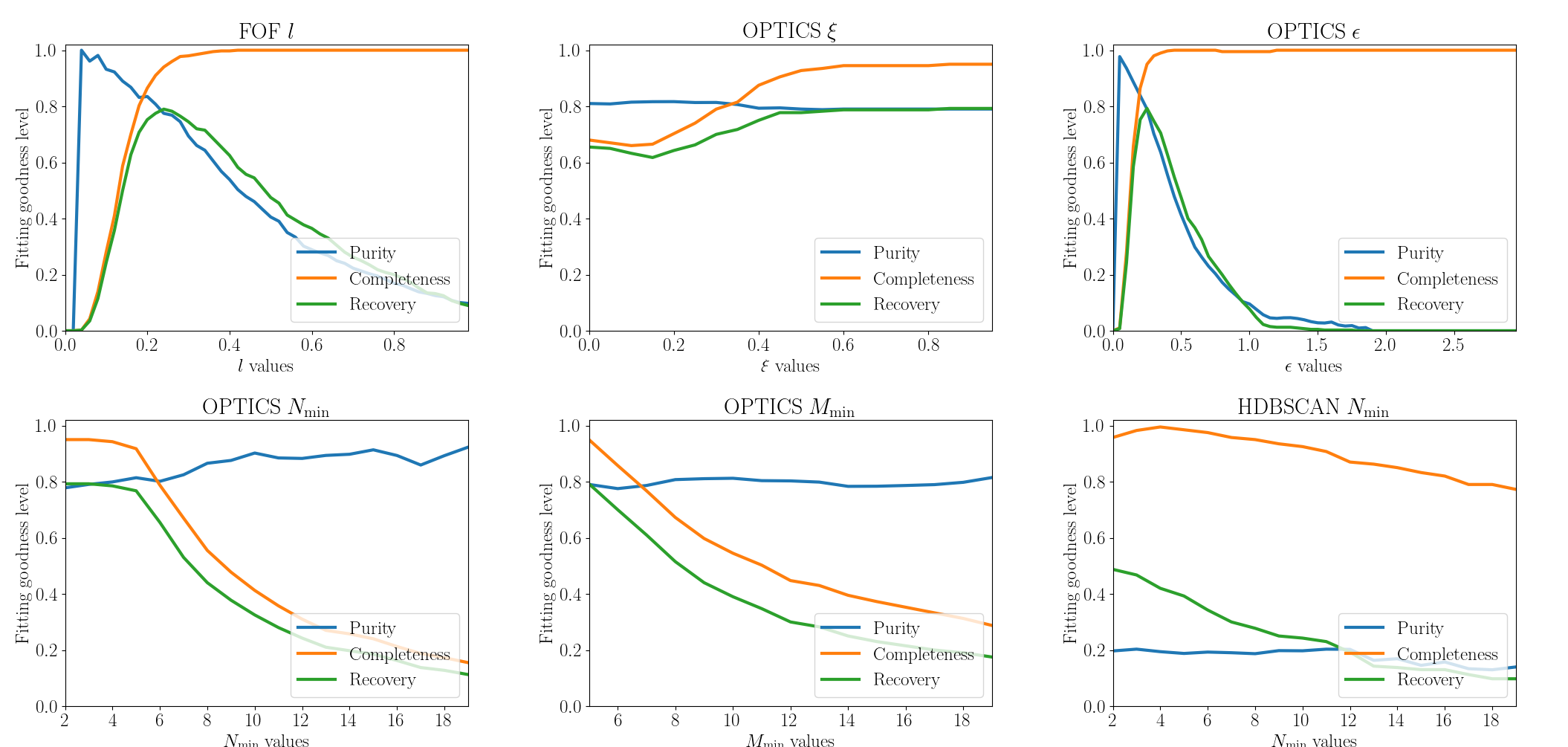}
    \caption{Results of parameter sensitivity tests conducted for the clustering algorithms FoF, HDBSCAN, and OPTICS, respectively. The metrics used to evaluate these tests were purity, completeness, and recovery rate, which were calculated by comparing the algorithms' outputs with group catalogs derived from the Millennium Simulation.}
    \label{fig:parameter_sensitivity_general}
\end{figure*}

Even though FoF and OPTICS are comparable in predicting galaxy groups, they differ significantly in their hyperparameter complexities. OPTICS requires tuning four hyperparameters, providing more flexibility and necessitating more careful configuration. On the other hand, FoF has only one primary hyperparameter, simplifying its use but potentially limiting its adaptability. This contrast raises questions about the sensitivity of their respective hyperparameters.
To investigate this, purity, completeness, and recovery rates were calculated for each algorithm under different values of a single hyperparameter while keeping the others unchanged at their respective optimal values. Figure \ref{fig:parameter_sensitivity_general} presents results based on soft criteria. 
These analyses were conducted on the same subsample described in Section \ref{sec:data-simulation}.

Our analysis reveals that the FoF algorithm exhibits significant sensitivity to the linking length parameter over this hyperparameter's entire possible value range. This dependency underscores the importance of careful tuning of the linking length parameter to ensure reliable identification of galaxy groups using the FoF method.

In comparison, the OPTICS results are primarily influenced by the minimum number of members $M_\mathrm{min}$ and minimum number of neighbors $N_\mathrm{min}$ parameters. Although the choice of $M_\mathrm{min}$ and $N_\mathrm{min}$ significantly affects the OPTICS results, it is noteworthy that setting these parameters to small values, such as in the range of 2 to 5, can achieve high completeness and purity in identifying galaxy groups and clusters. Increasing these parameters does not adversely affect the purity of the identified groups but only may reduce completeness. Consequently, choosing small values for $M_\mathrm{min}$ and $N_\mathrm{min}$ can be an appropriate strategy in the context of galaxy group and cluster identification, as it enables the algorithm to detect as many groups as possible from the entire data survey, including those with only a few members. Conversely, choosing larger values for $M_\mathrm{min}$ and $N_\mathrm{min}$ enables the focus on giant clusters, enhancing confidence in their identification.

As for the other two parameters of OPTICS, conventionally, the $\epsilon$ parameter primarily drives OPTICS results because, by definition, points lacking sufficient neighbors within an $\epsilon$-radius are classified as isolated, which is crucial for noise identification. However, surprisingly, the results exhibit remarkable stability for $\epsilon$ values exceeding 1.0, and even extreme choices still yield similar outcomes. This robustness can be attributed to how OPTICS extracts clusters from the reachability plot, where $\xi$ plays a dominant role. 
For $\epsilon$ values in a proper range, $\xi$ has less effect on clustering results, as depicted in Figures \ref{fig:parameter_sensitivity_general}.
This is because, in astronomical data, where galaxy groups and clusters are often more spatially distinct and less densely packed than objects in other types of datasets (like social networks or biological data), the natural separation between groups or clusters is already pronounced, reducing the need for fine-tuning $\xi$.

Finally, while HDBSCAN demonstrates low sensitivity to hyperparameter choices, its group prediction accuracy falls short, excluding its further consideration in this work.

In addition, it is noteworthy that extreme values of $\epsilon$ and $N_{\text{min}}$ in the OPTICS algorithm can achieve purity rates as high as 100\%. This feature of OPTICS highlights its capacity to precisely and effectively identify the densest regions within galaxy clusters. Furthermore, this insight indicates a new approach to locating BCGs efficiently. The detection and analysis of BCGs are crucial for understanding the mass distribution in clusters and the evolutionary dynamics involved. In Section \ref{sec:test-BCG}, we will explore the application of this method for BCG identification, evaluating its efficiency and broader implications.

\section{Test with real-world group catalog}
\label{sec:test-SDSS}

In Section \ref{sec:test-simulation}, we have demonstrated the efficacy of the OPTICS algorithm, particularly highlighting its stability in parameter sensitivity tests compared to the FoF method. Nonetheless, applying to real observational data remains a unique challenge not encountered in simulations. For instance, the number density distribution of astronomical objects is significantly constrained by the limitations inherent to telescopes and surveys, as well as by environmental factors and redshift variations. A particularly critical issue that cannot be overlooked is the redshift-space distortion, which introduces complexities not accounted for in simulation-based analyses. Various models have been proposed to investigate redshift-space distortions in galaxy surveys. These include the Eulerian dispersion model \citep{kaiserClusteringRealSpace1987}, the Lagrangian perturbation model \citep{buchertLagrangianTheoryGravitational1992, bouchetPerturbativeLagrangianApproach1995} and the Gaussian streaming model \citep{reidAccurateModelRedshiftspace2011, reidClusteringGalaxiesSDSSIII2012}, along with their variations. 
These models, including dispersion models and those expressing the redshift-space correlation function as an integral of the real-space correlation function, have been tested in configuration space to understand their predictive capabilities. It is shown that some models fitting simulations well over limited scales (on scales above $25-30\, h^{-1}$ Mpc) but failing at smaller scales \citep{whiteTestsRedshiftspaceDistortions2015}. 
This limitation poses challenges in accurately correcting the identification of galaxy groups and clusters, typically smaller in scale. The random velocities of galaxies in groups and clusters contribute significantly to redshift-space distortions on small scales, impacting the precision of these models in correcting for such distortions \citep{marulliRedshiftspaceDistortionsGalaxies2017}.

Consequently, extrapolating conclusions derived from simulations to real observational contexts requires caution. To address this, our research extends into the empirical evaluation of the FoF and OPTICS algorithms with real-world observational data of galaxies and galaxy groups, considering the effects of redshift-space distortions.

\subsection{Data Sample}
\label{sec:data-sdss}

To conduct the evaluation of FoF and OPTICS on real-world observations, we adopt data from the seventh Sloan Digital Sky Survey \citep[SDSS DR7;][]{abazajianSeventhDataRelease2009}. More specifically, we make use of the New York University Value-Added Galaxy Catalog \citep[NYU-VAGC;][]{blantonNewYorkUniversity2005}, which is based on SDSS DR7 but includes a set of significant improvements over the original pipelines. We select all galaxies in the main galaxy sample from this catalog using the identical selection criteria described in \citet{yangGalaxyGroupsSDSS2007}. This leaves 639,359 galaxies with reliable r-band magnitudes and measured redshifts from the SDSS DR7.

For our comparative analysis, we utilize the group and cluster catalog by \citet[][hereafter \citetalias{yangGalaxyGroupsSDSS2007}]{yangGalaxyGroupsSDSS2007}, updated to the version incorporating data from SDSS DR7 as a foundational reference. Among the three versions of group catalogs provided in \citetalias{yangGalaxyGroupsSDSS2007}, we adopt the one that is constructed using the SDSS model magnitude and includes additional SDSS galaxies with redshifts from alternative sources. The selection of the group centers, which are also BCG candidates, in this catalog is based on luminosity, as detailed by \citet{yangGalaxyOccupationStatistics2005} in Section 3.2. 

\subsection{Cure the Redshift-Space Distortion via sOPTICS with a LOS Scaling Factor}
\label{sec:scaled-optics}

\begin{figure}
    \centering
    \includegraphics[width=1\linewidth]{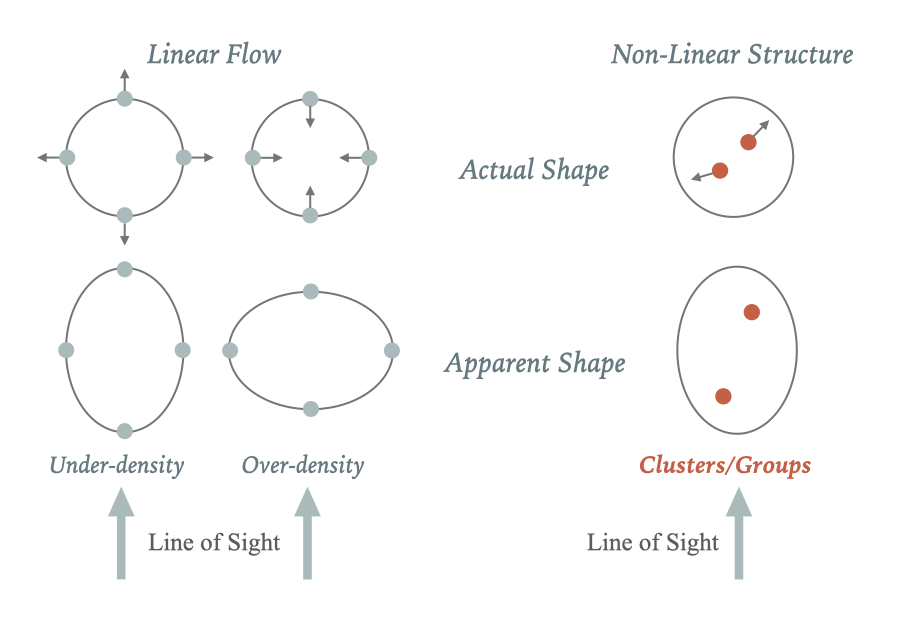}
    \caption{Illustration of redshift-space distortions: On the right, the figure illustrates the Kaiser effect, where on large scales, velocity flows into large overdensities compress the appearance of these structures along the line of sight. On the left, the figure illustrates the scenario on a smaller scale within galaxy groups and clusters. The random motions of galaxies within these compact environments result in an apparent elongation along the line of sight. This phenomenon is known as the "Fingers of God" effect.}
    \label{fig:distortion}
\end{figure}

As mentioned at the beginning of this section, one unavoidable challenge arises before applying OPTICS to real observations. Due to the redshift-space distortion phenomenon, galaxy groups exhibit an elongated appearance along the line of sight when observed in Cartesian coordinates (see Figure \ref{fig:distortion}). When applied in a three-dimensional space, this elongation presents a significant challenge for clustering algorithms, such as OPTICS. Specifically, it results in an underestimation of the true spatial extent of these groups. Consequently, galaxies relatively further away along the line of sight may be erroneously excluded from their respective groups. This misclassification can have notable implications for astrophysical studies, including inaccuracies in determining the centers of galaxy clusters and identifying BCGs. A careful consideration of the effects of redshift-space distortion is, therefore, vital in astrophysical cluster analysis to ensure the integrity and accuracy of the findings.

To address the issue of redshift-space distortion in clustering galaxy groups, we propose modifying the Euclidean distance metric typically employed in OPTICS clustering algorithms. This modification aims to counteract the elongation effect along the line of sight arising from redshift distortion. The adjustment involves scaling the distance calculation's line-of-sight (LOS) component.

The standard Euclidean distance between two points in a three-dimensional (3D) space is defined as:
\begin{equation}
D^2(u, v) = \sum_{i=1}^{3}(u_i - v_i)^2\,,
\end{equation}
where $u$ and $v$ represent the position vectors of the two points in space.

To address redshift-space distortion, we introduce a LOS scaling factor denoted as $s_\mathrm{LOS}$ for the line-of-sight component. 
This factor is applied specifically to the LOS component, thereby “shortening” distances in that direction and mitigating the effects of redshift-space distortion. We define the Elongated Euclidean Distance as:
\begin{equation}
D_\mathrm{Elongated}^2\bigl(u, v, s_\mathrm{LOS}\bigr)
= d_\mathrm{Transverse}^2(u, v) + d_\mathrm{LOS,,scaled}^2\bigl(u, v, s_\mathrm{LOS}\bigr).
\end{equation}
Here, $d_\mathrm{Transverse}(u, v)$ corresponds to the component of the Euclidean distance perpendicular to the LOS. And $d_\mathrm{LOS,\,scaled}$ is the scaled LOS component. In principle, one could define:
\begin{equation}
d_\mathrm{LOS}(u, v)
= \frac{\sum_{i=1}^{3} \bigl(u_i - v_i\bigr),u_i}{\sqrt{\sum_{i=1}^{3} \bigl(u_i\bigr)^2}} ,,
\end{equation}
and then multiply by $s_\mathrm{LOS}$ to obtain:
\begin{equation}
d_\mathrm{LOS,,scaled}\bigl(u, v, s_{\mathrm{LOS}}\bigr)
= s_{\mathrm{LOS}} d_\mathrm{LOS}(u, v).
\end{equation}

However, since $d_\mathrm{LOS}(u, v)$ does not equal $d_\mathrm{LOS}(v, u)$, the resulting distance may lose symmetry. In other words, $D_\mathrm{Elongated}(u, v) \neq D_\mathrm{Elongated}(v, u)$, which can undermine the metric properties typically assumed by OPTICS and potentially increase the algorithm’s sensitivity to data ordering. 
To ensure a symmetric distance measure and preserve the stability of core-point definitions, we adopt a symmetrized version of the LOS component:
\begin{equation}
d_\mathrm{LOS,,scaled}^\mathrm{sym}\bigl(u, v, s_{\mathrm{LOS}}\bigr)
= s_{\mathrm{LOS}} \frac{d_\mathrm{LOS}(u, v) + d_\mathrm{LOS}(v, u)}{2}.
\end{equation}
This modification guarantees that $d_\mathrm{LOS,\,scaled}^\mathrm{sym}(u, v, s_\mathrm{LOS}) = d_\mathrm{LOS,\,scaled}^\mathrm{sym}(v, u, s_\mathrm{LOS})$, thereby restoring symmetry to the overall distance function. By enforcing such symmetry, one retains the theoretical benefits of a core-distance-based reachability measure in OPTICS—namely, stable cluster structures that are less dependent on the processing order of data points—and ensures that the clustering results remain interpretable as a function of an actual distance metric.

\begin{figure}
    \centering
    \includegraphics[width=1\linewidth]{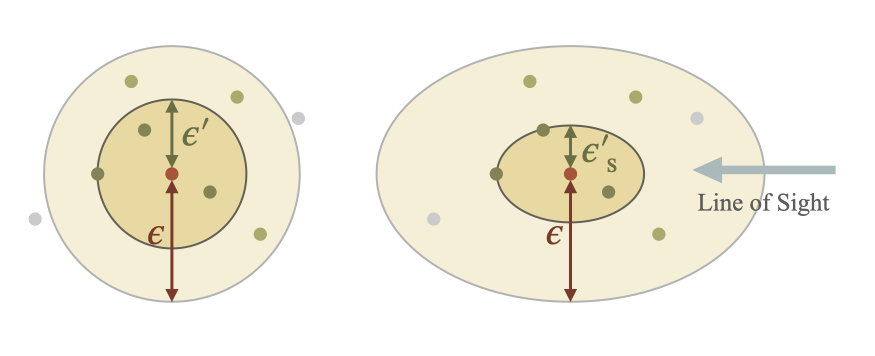}
    \caption{Illustration demonstrating the role of the LOS scaling factor $s_\mathrm{LOS}$ in our sOPTICS algorithm to mitigate redshift-space distortion.}
    \label{fig:scaling_los}
\end{figure}

Figure~\ref{fig:scaling_los} illustrates the elongated Euclidean distance's effect on the OPTICS clustering results. This adjustment transforms the sphere with an $\epsilon$-radius into an ellipse elongated along the LOS, enabling the inclusion of more distant galaxies along the LOS as possible neighbors. Consequently, this results in shorter core distances and deeper valleys in the reachability plot. Unlike direct modeling of redshift dispersion, this approach indirectly addresses and mitigates the underestimation issues associated with redshift-space distortions.

\begin{figure*}
    \centering
    \includegraphics[width=0.33\textwidth]{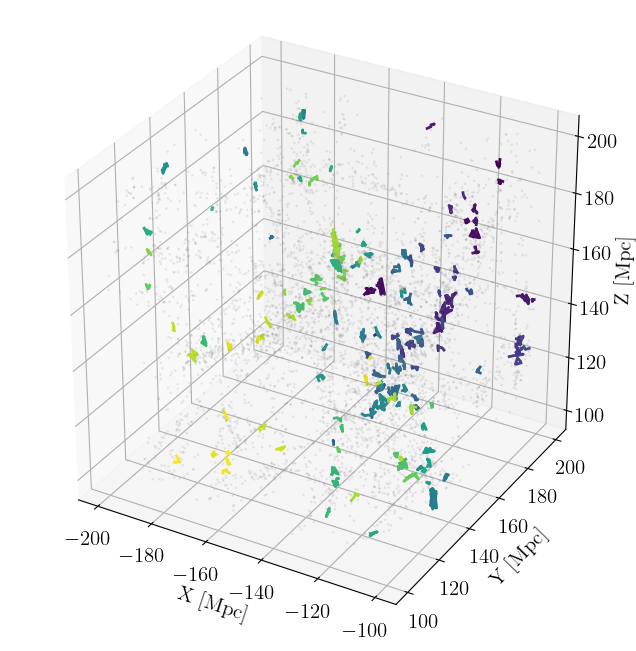}
    \includegraphics[width=0.33\textwidth]{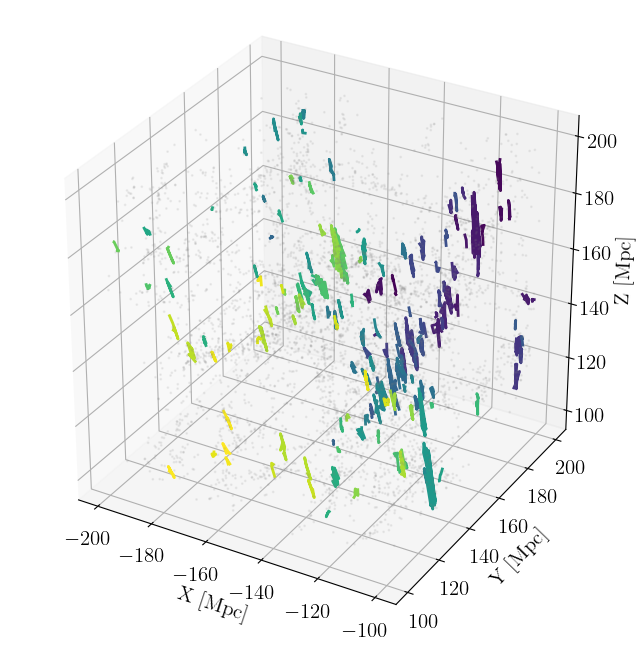}
    \includegraphics[width=0.33\textwidth]{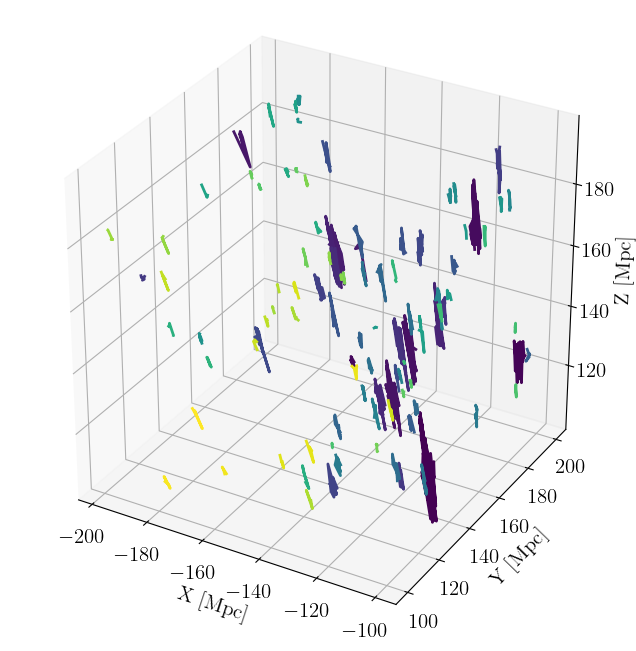}
    \caption{The left panel of the figure presents the result of the sOPTICS algorithm applied without any scaling adjustments. The middle panel presents the results of unscaled OPTICS, while the pairs of LOS distances have been scaled by a factor of $s_\mathrm{LOS} = 0.2$. The right panel, in contrast, displays the actual galaxy groups as identified in the \citetalias{yangGalaxyGroupsSDSS2007} group catalog. In the left and middle panels of the figure, the gray points represent galaxies that the OPTICS and sOPTICS algorithm predicts as not belonging to any groups.}
    \label{fig:OPTICS-w/o-scaling}
\end{figure*}

In practice, we have tested the effectiveness of this scaling adjustment. Figure \ref{fig:OPTICS-w/o-scaling} shows the clustering results of OPTICS in a subsample at redshift $z=0.10$, both with and without the scaling adjustment, and compares the predicted groups to the \citetalias{yangGalaxyGroupsSDSS2007} groups. The results demonstrate that redshift-space distortion significantly influences clustering outcomes, leading to a considerable underestimation along the LOS. By employing the elongated Euclidean distance, we have achieved a more precise prediction of galaxy groups, which improves both the detection of group shapes and the accuracy of membership. In the subsequent sections, we will refer to this OPTICS clustering method with the elongated Euclidean distance as scaled OPTICS (sOPTICS).

\begin{figure*}
    \centering
    \includegraphics[width=0.85\textwidth]{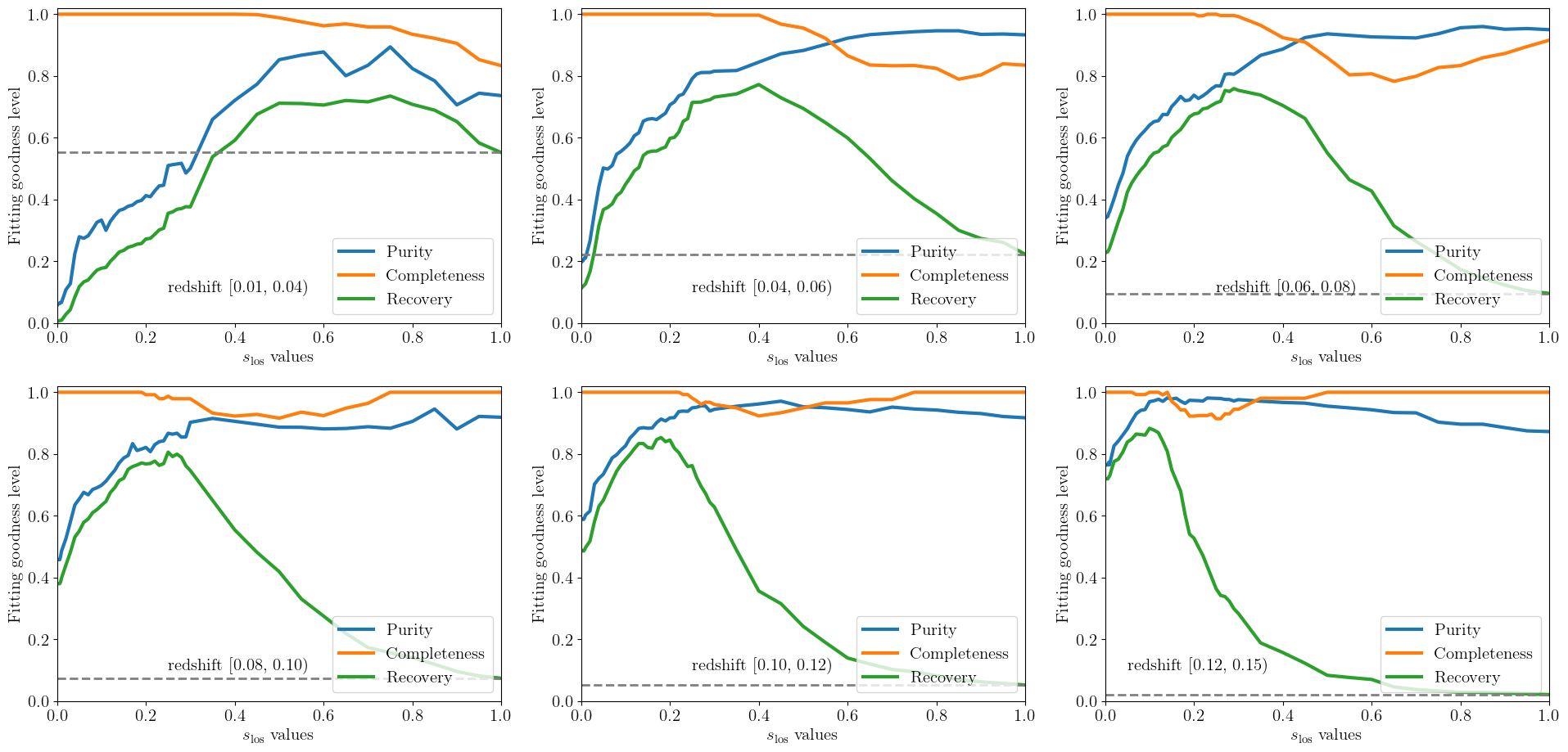}
    \caption{The dependence of purity, completeness, and recovery rate on the LOS scaling factor $s_\mathrm{LOS}$ across different redshifts for a specific sample defined by a right ascension range between 150 and 200 deg and a declination range between 10 and 60 deg. The grey dashed line highlights the baseline of recovery rate achieved by the original OPTICS method when $s_\mathrm{LOS}=1$.}
    \label{fig:Scale_los-redshift}
\end{figure*}

However, the effect of redshift-space distortion is not constant across different redshifts. A galaxy with a cosmological redshift $z_{\rm c}$ and a 'peculiar' redshift $z_{\mathrm{p}}$ will appear to an observer to have an observed redshift $z$, as described by the equation:
\begin{equation}
(1+z) = (1+z_{\rm c})\left(1+z_{\mathrm{p}}\right)\;.
\end{equation}
The approximation $z= z_{\rm c}+ z_{\mathrm{p}}$ is only valid for small redshifts. Consequently, redshift-space distortion increasingly affects galaxy groups at higher redshifts. This stronger distortion necessitates more robust adjustments, specifically, smaller $s_\mathrm{LOS}$. Therefore, it is necessary to adjust the value of the LOS scaling factor $s_\mathrm{LOS}$ with redshift.
To ascertain the proper values of $s_\mathrm{LOS}$ correcting for redshift-space distortions across varying redshift bins, in Section \ref{sec:parameter} we iteratively optimized $s_\mathrm{LOS}$ by maximizing the concordance between our results and the \citetalias{yangGalaxyGroupsSDSS2007} group catalog.
Figure \ref{fig:Scale_los-redshift} illustrates how the optimal LOS scaling factor changes with redshift. It is clearly shown that the optimal LOS scaling factor decreases with higher redshifts, indicating that the Euclidean distance along the LOS is elongated more significantly. This trend is consistent with theoretical predictions.

In addition, the LOS scaling factor is also related to the values of $\epsilon$ itself. If the $\epsilon$ value is sufficiently large, all potential group members would be included, eliminating the need for elongation along the LOS. 
However, the redshift-space distortion predominantly impacts the distance measurements along the LOS. As Figure \ref{fig:parameter_sensitivity_general} demonstrates, employing larger $\epsilon$ values might reduce the identified groups' purity. This reduction in purity occurs because a larger $\epsilon$ value causes the algorithm to excessively consider neighboring objects along the LOS and in the opposite direction, with little physical association.
Considering the inherent noise and variability in the observed distribution of galaxies compared to their simulated counterparts, carefully adjusting the parameters $\epsilon$ and the corresponding $s_\mathrm{LOS}$ is essential. To address this, we have examined the relationship between the optimal sets of $s_\mathrm{LOS}$ and $\epsilon$. Our findings, detailed in Figure \ref{fig:two-param-fit2}, reveal a well-defined optimal region for selecting these parameters. This optimal region ensures a balanced approach to grouping galaxies, optimizing both the purity of the groups and the inclusion of genuine group members, thus mitigating the effects of observational noise and distortion.

\subsection{Choices of Hyperparameter Values}
\label{sec:parameter}

To determine the optimal hyperparameter values for FoF and sOPTICS, similar to our approach for refining $s_{\mathrm{LOS}}$, we initiate the optimization process by aligning them with the \citetalias{yangGalaxyGroupsSDSS2007} group catalog, which serves as our reference model. Incorporating the LOS scaling factor $s_{\mathrm{LOS}}$ into sOPTICS, 
the algorithm now boasts five hyperparameters requiring optimization, whereas FoF requires only two: linking length $l$ and $M_{\rm min}$. 
It is important to note that the evaluation criteria diverge from the tests conducted on simulated galaxy catalogs as described in Section \ref{sec:test-simulation}. This divergence stems from astrophysical studies on galaxy groups and clusters typically prioritize those with substantial membership. Given that merely 1.78\% of groups consist of at least five galaxy members (totaling 8,427 out of 472,416 groups in \citetalias{yangGalaxyGroupsSDSS2007} catalog), we suggest a refined adjustment to the definition of the recovery rate, strategically assigning heightened weight to groups exhibiting a greater abundance of members:
\begin{equation}
    F_{\mathrm{R}} = 
    \sum^{\mathcal{N}}_{i=0} \delta_i \times \frac{ \text{Number of galaxies in group $i$}}
    {\text{Total number of non-isolated galaxies}}\, ,
\end{equation}
where $\mathcal{N}$ is the total number of true groups in \citetalias{yangGalaxyGroupsSDSS2007}, and:
\begin{equation}
    \delta_i = 
\begin{cases}
1,  & \text{if group $i$ is simultaneously pure and complete,} \\
0, & \text{otherwise.}
\end{cases}
\end{equation}
Leveraging the abundance-weighted recovery rate as a criterion for optimization allows us to prioritize identifying giant clusters in our analyses. When comparing different sets of parameters, preference is given to those configurations that enhance the recovery of a larger number of giant clusters, as cataloged in \citetalias{yangGalaxyGroupsSDSS2007}.

To fine-tune the hyperparameters, we select ten subsamples from low redshift galaxies ($z<0.05$), each with a cubic side length of 100 Mpc. Then, we first identify the optimal values of hyperparameters of FoF, as well as sOPTICS with a constant $s_\mathrm{LOS}$. 
The search spaces for these hyperparameters are identical to that listed in Table \ref{tab:hypervalues}, and the optimal values are detailed in Table \ref{tab:hypervalue-optimal}.
Using these hyperparameters, we achieved a maximum recovery rate of $F_\mathrm R=0.8$ for FoF and 0.76 for sOPTICS. It is important to note that for sOPTICS, although the recovery rate of 0.76 wasn't the peak for every individual test subsample—with the highest rate reaching 0.89 in certain scenarios—these hyperparameters yield the most consistent and accurate predictions of BCGs across the board, as elaborated in Section \ref{sec:test-BCG}. Thus, we adopted this set of hyperparameters for sOPTICS as the most suitable choice, balancing overall performance across various testing conditions. 

Maintaining the optimal hyperparameters identified earlier, we proceeded to select subsamples of 100 Mpc within each redshift bin to fit the optimal value for $s_\mathrm{LOS}$ that effectively mitigates redshift space distortion, as detailed in Table \ref{tab:hypervalue-s_los}. Here we note that, after extensive testing, we meticulously determined the delineation of redshift bins with prior knowledge from the \citetalias{yangGalaxyGroupsSDSS2007} group catalog to prevent the segmentation of giant clusters across two bins, ensuring a more coherent and accurate analysis.

{
\renewcommand{\arraystretch}{1.2}
\begin{table}
    \centering
    \begin{tabular}{ccccccccc}
    \hline
        \multicolumn{2}{c}{FoF} & & & & \multicolumn{4}{c}{sOPTICS} \\
    \hline
        $l$ & $M_\mathrm{min}$ & & & & $\epsilon$  &   $N_\mathrm{min}$ &  $M_\mathrm{min}$   & $\xi$ \\
        1.8 & 5 & & & & 1.2    & 5  &   5   &   0.9 \\
    \hline
    \end{tabular}
    \caption{The hyperparameters for the FoF and sOPTICS algorithm, as applied to the entire galaxy sample described in Section \ref{sec:data-sdss}. It is important to note that, due to the magnitude limit constraints of the SDSS survey and local incompleteness factors, the values of $N_\mathrm{min}$ were adjusted based on the redshifts of the galaxies. For galaxies with redshifts $z<0.10$, we utilized the hyperparameters as listed, and for galaxies at higher redshifts, we modified the $N_\mathrm{min}$ to 4.}
    \label{tab:hypervalue-optimal}
\end{table}

\begin{table}
    \centering
    \begin{tabular}{ccccc}
    \hline
    Redshift bins &  0.01 - 0.04 & 0.04 - 0.06 & 0.06 - 0.08 & 0.08 - 0.10 \\
    $s_\mathrm{LOS}$  &  0.5 & 0.4 & 0.35 & 0.3 \\
    \hline
    Redshift bins & 0.10 - 0.12 & 0.12 - 0.15 & 0.15 - 0.20 & \\
    $s_\mathrm{LOS}$ & 0.19 & 0.08 & 0.01 &   \\
    \hline
    \end{tabular}
    \caption{The best-fitted values of the LOS scaling factor, $s_\mathrm{LOS}$, across various redshifts. The analysis is constrained by other parameters as listed in Table \ref{tab:hypervalue-optimal}. It is important to note that the bin settings provided here represent approximate values. In our practice, slight adjustments are made to the settings of redshift bins to prevent significant groups from being split across two bins.}
    \label{tab:hypervalue-s_los}
\end{table}
}

\subsection{Basic Results}

With the optimal values for hyperparameters and the LOS scaling factor listed in Table \ref{tab:hypervalue-optimal} and \ref{tab:hypervalue-s_los}, the overall abundance-weighted recovery rate for galaxy groups using the sOPTICS algorithm is 75.0\%, with the abundance-weighted purity of 86.6\% and completeness of 97.1\%. The total number of identified groups is 12,242, while there are 8,427 groups with at least 5 member galaxies in \citetalias{yangGalaxyGroupsSDSS2007} catalog. Meanwhile, the soft recovery rate stands at 69.2\%, indicating that we can precisely identify 5,831 galaxy groups, with two-thirds of their member galaxies matching those of the true groups identified in \citetalias{yangGalaxyGroupsSDSS2007} and covering more than half of the actual members, out of a total of 8,427 true groups. In contrast, the FoF algorithm achieved a soft recovery rate of 42.6\%, while with a purity of 44.2\% and completeness of 54.7\% The total number of identified groups using FoF is 12,396. Therefore, by incorporating the LOS scaling factor, we significantly enhanced the precision of the sOPTICS algorithm's predictions. Consequently, in identifying galaxy groups and clusters, sOPTICS performs comparably to, and in some cases even better than, FoF when parameters are tuned based on sub-samples.

\begin{figure}
    \centering
    \includegraphics[width=0.9\linewidth]{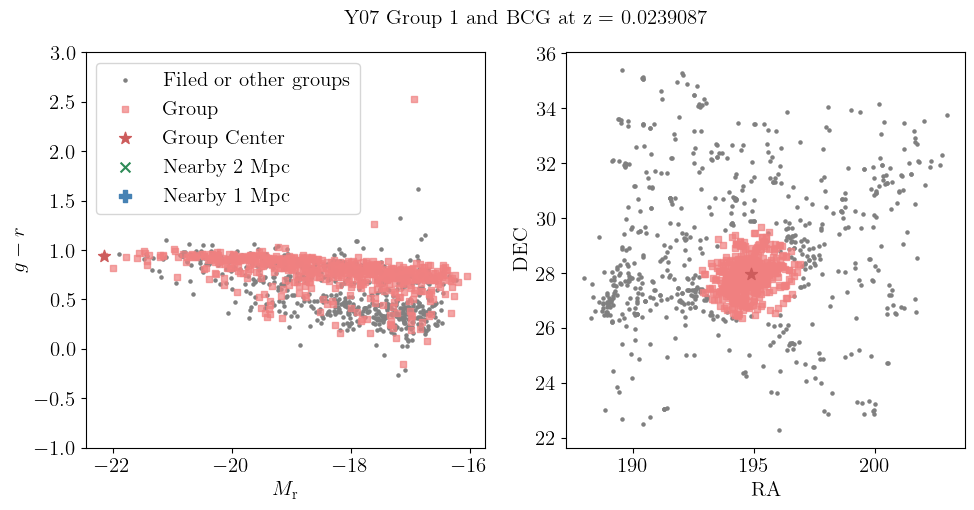}
    \includegraphics[width=0.9\linewidth]{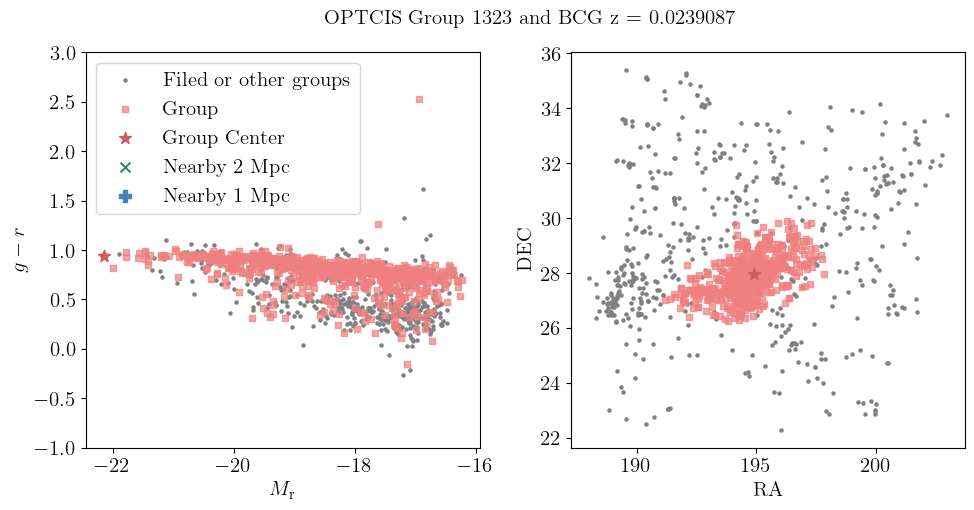}
    \caption{Color–magnitude diagrams (left panels) and projected distributions (right panels) of SDSS galaxies. The upper panels present galaxies in the largest group from the \citetalias{yangGalaxyGroupsSDSS2007} catalog (group ID 1) and field galaxies surrounding the group center within a 10 Mpc radius. The bottom panels present the corresponding group predicted by sOPTICS and its surrounding field galaxies. These color-magnitude diagrams reveal a tight correlation in the color-magnitude space, closely resembling a cluster red sequence. Meanwhile, the original group from the \citetalias{yangGalaxyGroupsSDSS2007} catalog is effectively predicted by sOPTICS, exhibiting high completeness, and the BCG of this group is precisely identified.}
    \label{fig:CMdiagram}
\end{figure}

We also visually inspect the large galaxy groups and clusters with the aid of the color-magnitude relation of the groups and clusters, $r$-magnitude versus $(g - r)$ Color diagrams were made for field galaxies and cluster + field galaxies in each SDSS square. The stacked field galaxy maps were subtracted from the stacked cluster galaxy maps, taking into account the relative areas (within 10 Mpc). In the color-magnitude diagram, the presence of a clear red sequence indicates a real galaxy group, as opposed to just a chance alignment of field galaxies. Figure \ref{fig:CMdiagram} shows the color–magnitude diagrams for the largest galaxy group in \citetalias{yangGalaxyGroupsSDSS2007} and the corresponding group predicted by sOPTICS, which is also the largest one in prediction. The maps reveal a distinct trend in color-magnitude space, resembling a cluster red sequence. Remarkably, over two-thirds of the member galaxies, including the BCG, are accurately predicted.

\section{Capability of Clustering Algorithms}
\label{sec:discussion}

With the aid of the LOS scaling factor, we have successfully recovered nearly 70~\% of galaxy groups from the \citetalias{yangGalaxyGroupsSDSS2007} group catalog. This is a significant improvement considering the complexity and time-intensive nature of the \citetalias{yangGalaxyGroupsSDSS2007} catalog identification process.

In \citealt{yangGalaxyGroupsSDSS2007}, initially, the FoF algorithm with very short linking lengths in redshift space was used to identify preliminary groups that likely represent the central regions of these clusters. The geometrically determined, luminosity-weighted centers of all FoF-identified groups with at least two members were designated potential group centers. Galaxies not associated with these FoF groups were also treated as potential centers. Each group’s characteristic luminosity, \(L_{19.5}\), was then calculated to facilitate a meaningful group comparison. This luminosity was used to assign a halo mass to each group, which allowed for the estimation of the group’s halo radius and velocity dispersion. Subsequent updates to group memberships were guided by a probability density function calculated in redshift space around each group’s center, considering halo properties. This iterative process -- consisting of updating group memberships, recalculating centers, and refining the \(M_h/L_{19.5}\) to \(L_{19.5}\) relationship -- continued until the group dynamics stabilized, usually after a few iterations. Their comprehensive method thus enhanced the understanding of galaxy group dynamics and composition, overcoming limitations posed by redshift space distortions.

In comparison, our method scaled OPTICS, takes only about one hour on average and involves a straightforward process, yet it achieves high recovery rates of the \citetalias{yangGalaxyGroupsSDSS2007} catalog. Therefore, the primary strength of sOPTICS lies in its efficiency in identifying galaxy groups from large surveys with very low computational costs. Moreover, it is particularly sensitive to detecting large clusters, achieving high accuracy in identifying their members. This combination of speed, simplicity, and precision makes sOPTICS an advantageous tool for astrophysical studies requiring the analysis of extensive data sets.

\subsection{Interdependence of Hyperparameters in sOPTICS}

\label{sec:discussion-OPTICS}

\begin{figure}
    \centering
    \includegraphics[width=1\linewidth]{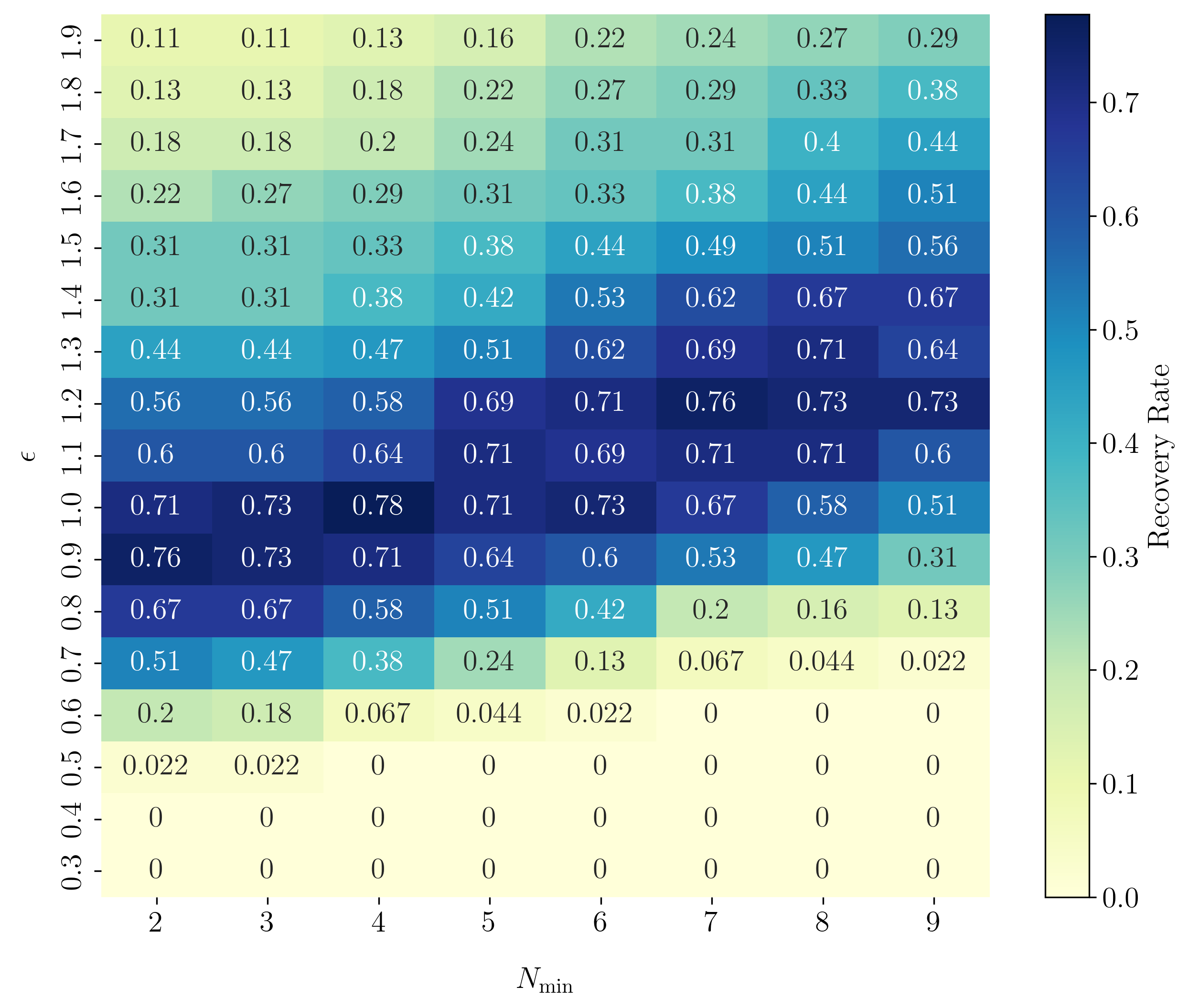}
    \caption{The results of galaxy group finding for various pairs of values of $\epsilon$ and $N_\mathrm{min}$. The color bar represents the weighted recovery rate of the predicted groups relative to a subsample from the \citetalias{yangGalaxyGroupsSDSS2007} groups catalog, covering a $30^2 \, \deg ^2$  sky area within the redshift range $0.7<z<0.8$.}
    \label{fig:two-param-fit}
\end{figure}

\begin{figure}
    \centering
    \includegraphics[width=1\linewidth]{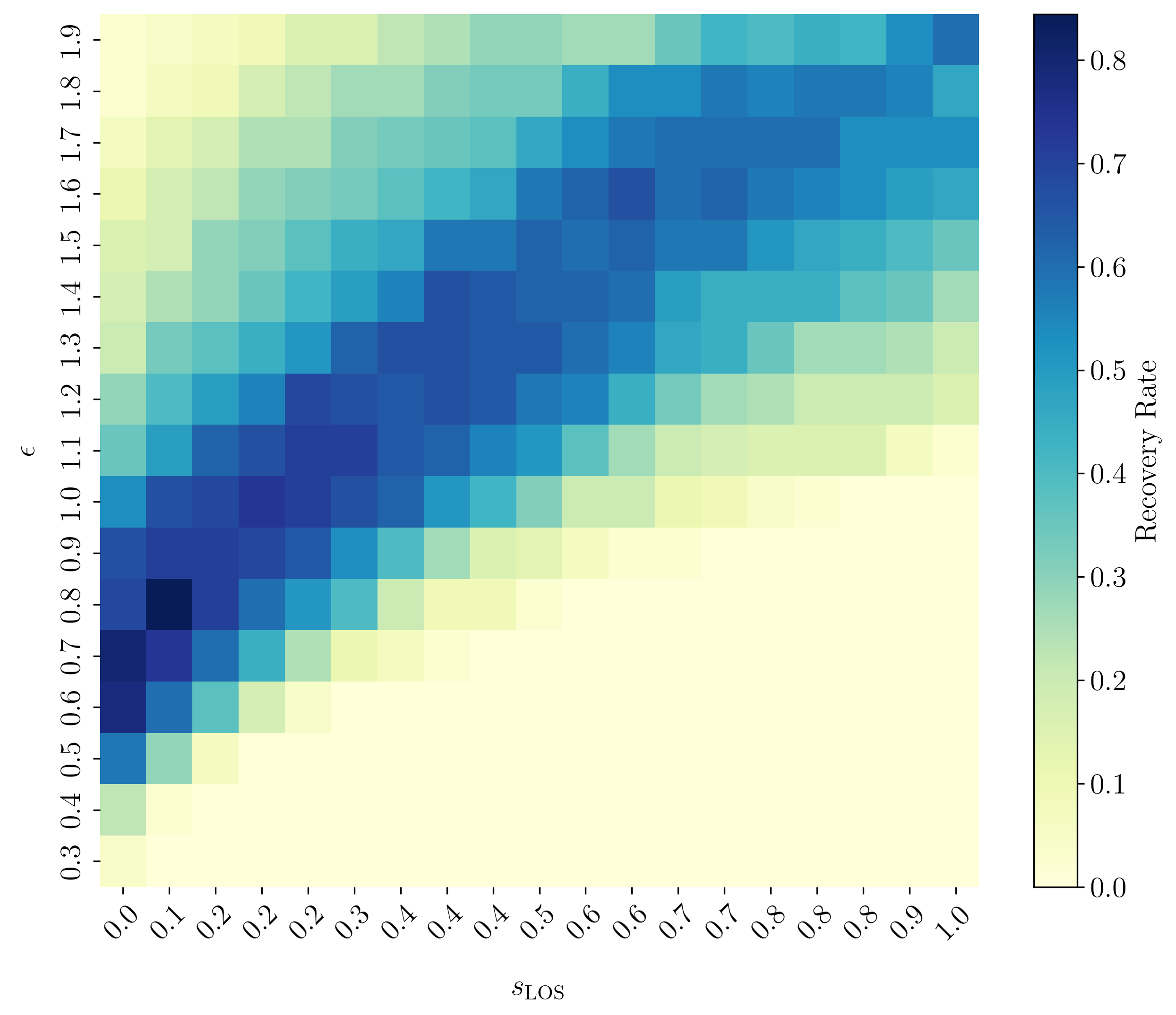}
    \caption{As Fig. \ref{fig:two-param-fit}, but showing the results of galaxy group finding for various pairs of values of $s_\mathrm{LOS}$ and $\epsilon$.}
    \label{fig:two-param-fit2}
\end{figure}

In employing the scaled OPTICS clustering algorithm to identify galaxy clusters, the hyperparameters $\epsilon$ (the maximum radius for neighborhood density estimation), $N_\mathrm{min}$ (the minimum number of points required to form a cluster), and the LOS scaling factor crucially influence the results, as detailed in Section \ref{sec:test-simulation}. These parameters are pivotal in determining reachability distances and adjusting the algorithm to mitigate the effects of redshift-space distortion.
Although one might anticipate a high sensitivity to parameter variations, sOPTICS exhibits resilience by maintaining an optimal range for these values. This finding is illustrated in Figures \ref{fig:two-param-fit} and \ref{fig:two-param-fit2}, where the interdependence of $\epsilon$ and $N_\mathrm{min}$, as well as $\epsilon$ and $s_\mathrm{LOS}$, is presented. Notably, a clear correlation emerges between $\epsilon$ and $N_\mathrm{min}$; as $\epsilon$ increases, $N_\mathrm{min}$ must also to be adjusted upward to maintain effective clustering. Essentially, to preserve the purity of the clusters identified by the sOPTICS algorithm, the criteria must shift toward identifying denser and larger clusters as the $\epsilon$ threshold is raised. The positive correlation suggests that $\epsilon^3 \propto N_\mathrm{min}$ aligns with theoretical expectations. Theoretically, this adjustment ensures that the increase in the neighborhood radius does not lead to the inclusion of outlier points or less dense areas. Thus $N_\mathrm{min}$ should correlate with the volume of space encompassed within $\epsilon$, which implies a cubic relationship ($\epsilon^3$).

Adjustments in the LOS scaling factor, $s_\mathrm{LOS}$, which modifies how the LOS distance is shortened, exhibit a nearly linear relationship with $\epsilon$, such that $\epsilon \propto s_\mathrm{LOS}$. This relationship implies that increasing $\epsilon$ expands the effective search radius in the clustering algorithm, thereby capturing more of the spatial distribution of galaxies affected by redshift space distortion. Consequently, it reduces the necessity to stretch the LOS distance to mitigate these distortions.

These relationships underscore the interconnected nature of $\epsilon$, $N_\mathrm{min}$, and $s_\mathrm{LOS}$, as well as their collective impact on optimizing cluster detection and recovery rates.
Specifically, selecting parameter values within this optimal range can yield galaxy groups with a high recovery rate comparable to those obtained from reliable group catalogs that require complex and computationally intensive processes. The optimal ranges depicted in Figures \ref{fig:two-param-fit} and \ref{fig:two-param-fit2} also identify a potential characteristic number density for categorizing galaxies in a survey as a group. Given the similar local completeness of the survey, this characteristic number density of galaxy groups can be applied to other observations. 

However, given the current redshift range of the observed data, the relationships observed between $\epsilon$, $N_\mathrm{min}$, and $s_\mathrm{LOS}$ are preliminary and roughly empirical. To more precisely define these relationships and understand the characteristic number density, a comprehensive analysis using both real-world data and mock catalogs is crucial. This approach would help determine whether the observed linear trend between $\epsilon$ and $s_\mathrm{LOS}$ is an artifact of the specific dataset used in this study or if it reflects a more general characteristic applicable across different galaxy cluster distributions. 

\subsection{Application of sOPTICS to an Independent Catalog}
\label{sec:Shi}

Beyond examining empirical parameter relationships in a controlled setting, a core advantage of sOPTICS is its adaptability to other observational datasets with minimal re-tuning of parameters. To explore this aspect, we applied sOPTICS to an independent catalog from \citet{shiMappingRealspaceDistributions2016}, which contains 586,025 galaxies from SDSS DR13 \citep{albareti13thDataRelease2017} spanning a redshift range of $0\leq z \leq 0.20$. The group catalog was constructed using the adaptive halo-based group finder developed by \citet{yangGalaxyOccupationStatistics2005, yangGalaxyGroupsSDSS2007}. Specifically, they corrected redshift-space distortions by reconstructing the large-scale velocity field to account for the Kaiser effect and statistically redistributing galaxies within halos based on an NFW profile to address the Finger of God effect, yielding a pseudo-real-space galaxy group catalog of 8,640 systems (each containing at least five member galaxies).

We adopt the same sOPTICS hyperparameters determined in Tables \ref{tab:hypervalue-optimal} and \ref{tab:hypervalue-s_los}, without additional fine-tuning. Under these settings, sOPTICS identified 10,057 predicted groups, achieving a recovery rate of 55.2\%, a purity of 63.7\%, and a completeness of 75.8\%. These results are notable because they demonstrate that sOPTICS’s parameter choices, optimized on a different dataset, remain effective when transferred to new observations.

In contrast, we apply the FoF linking length obtained from Tables \ref{tab:hypervalue-optimal} to the same catalog. This yield significantly lower performance: 9,421 predicted groups with a recovery rate of 30.8\%, a purity of 46.7\%, and a completeness of 50.2\%. The substantial disparity suggests that FoF is more sensitive to its linking length parameter, whereas sOPTICS is comparatively robust to modest changes in data characteristics. By adapting cluster boundaries based on local density structures, sOPTICS retains its effectiveness across multiple samples, reducing the need for exhaustive parameter searches.

\begin{figure}
    \centering
    \includegraphics[width=1\linewidth]{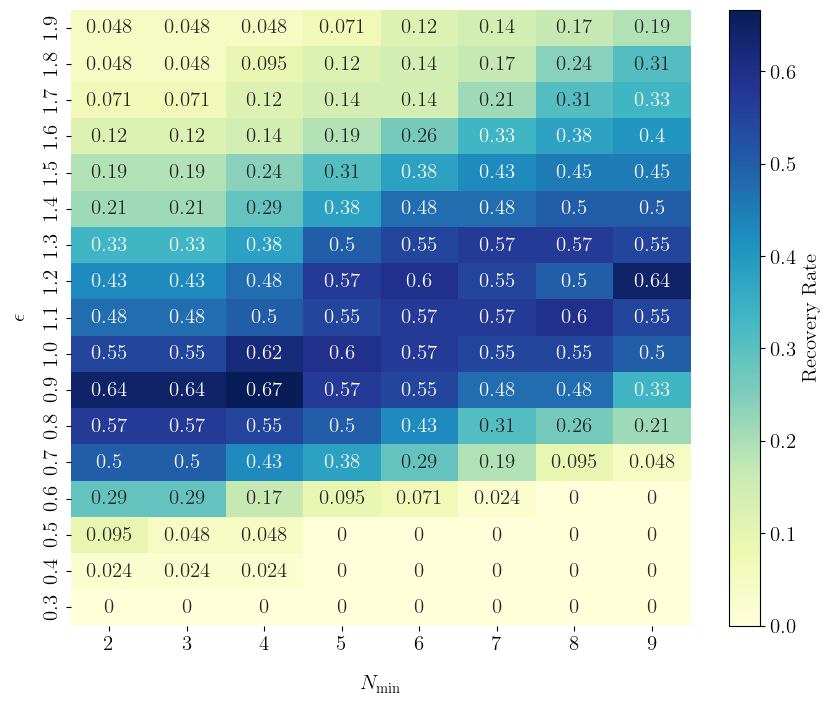}
    \caption{As Fig. \ref{fig:two-param-fit}, but showing the validation results of the predicted groups relative to a subsample from the groups catalog of \citet{shiMappingRealspaceDistributions2016}.}
    \label{fig:two-param-fit-shi}
\end{figure}

To investigate whether performance could be enhanced further, we examined the interdependence of $\epsilon$ and $N_\mathrm{min}$ on this new dataset while keeping other sOPTICS parameters unchanged (as in Figure \ref{fig:two-param-fit}). We find the same correlation pattern between these two parameters that was observed in our earlier tests, indicating that increasing $\epsilon$ necessitates higher values of $N_\mathrm{min}$ to preserve cluster purity. By simply optimizing $\epsilon$ and $N_\mathrm{min}$ alone, the recovery rate improved to 67\%, as presented in Figure \ref{fig:two-param-fit-shi}. This outcome underscores both the flexibility of sOPTICS and its potential for achieving higher cluster identification accuracy through modest parameter adjustments.

\subsection{Performance of finding BCGs}
\label{sec:test-BCG}

As demonstrated, sOPTICS can effectively detect large clusters with precise member identification, including the BCGs. To evaluate the performance of our sOPTICS method in identifying BCGs from a galaxy survey, we conducted a comparative test against a recent BCGs sample from \citet[][hereafter \citetalias{hsuSDSSIVMaNGACannibalism2022}]{hsuSDSSIVMaNGACannibalism2022}. Their parent BCG sample of 4,033 galaxy clusters is also extracted from the group catalog of \citetalias{yangGalaxyGroupsSDSS2007} with applying a cut in the cluster mass $M_{180 m} \geqslant 10^{14} h^{-1} M_{\odot}$. By cross-matching BCG candidates with the 8,113 galaxies released in the ninth Product Launch (MPL-9) of the Mapping Nearby Galaxies at Apache Point Observatory \citep[MaNGA][]{bundyOVERVIEWSDSSIVMaNGA2014}, they identified 128 BCGs situated within a redshift range of $z = 0.02 - 0.15$. These clusters are all detected in the X-rays by \citet{wangMeasuringXrayLuminosities2014}, which provides a cluster catalog with X-ray luminosity from the ROSAT All Sky Survey. However, \citetalias{yangGalaxyGroupsSDSS2007} primarily select BCGs based on luminosity, occasionally resulting in the selection of spiral galaxies as BCG candidates. Therefore, \citetalias{hsuSDSSIVMaNGACannibalism2022} implemented an additional visual selection process: if the BCG candidates show a spiral morphology or do not represent the most luminous galaxy on the red sequence, alternative candidates would considered. The cluster would be excluded if no superior candidate exists or MaNGA has not observed the more suitable candidate. As a result, 121 BCGs have been visually confirmed, of which 118 were originally part of the \citetalias{yangGalaxyGroupsSDSS2007} catalog.

While traditionally thought to be close to the center, recent studies have shown that the BCG may not always be at the cluster's center, with a fraction of BCGs being non-central depending on the halo mass \citep{chuPhysicalPropertiesBrightest2021}. This deviation from being at the center is due to different definitions of BCGs based on their luminosity or mass, regardless of their position within the cluster. Therefore, in this work, we identify BCGs based solely on their $r$ band magnitude, irrespective of their spatial position in the cluster.

\begin{figure*}
    \centering
    \includegraphics[width=0.45\textwidth]{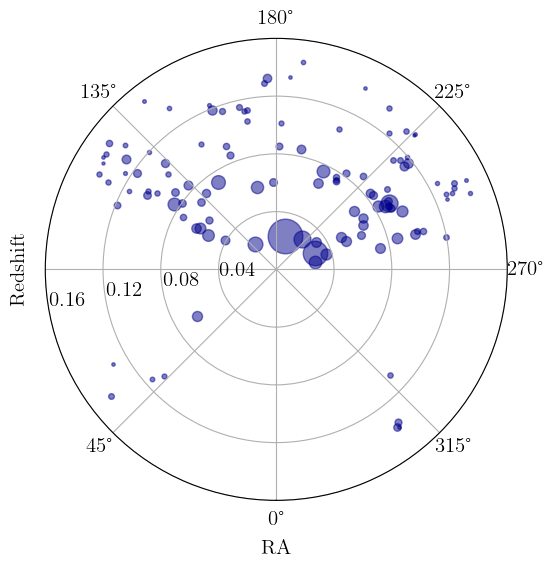}
    \includegraphics[width=0.45\textwidth]{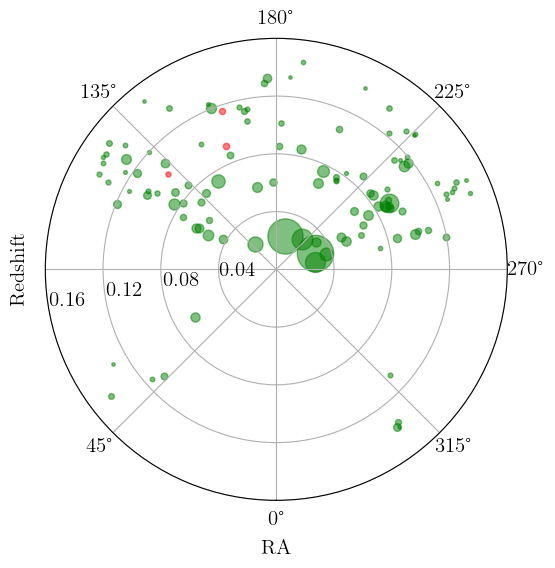}
    \caption{Distribution of both recovered and unrecovered BCGs from \citetalias{hsuSDSSIVMaNGACannibalism2022} within their respective galaxy groups visualized in the RA-redshift space. The left panel shows the distribution of galaxy groups in \citetalias{yangGalaxyGroupsSDSS2007} catalog. In the right panel, green dots represent the galaxy groups successfully identified by sOPTICS. And red dots represent the groups that remained undetected by sOPTICS. The size of these dots is scaled to correspond to the member density of each group.}
    \label{fig:BCG-distribution}
\end{figure*}

\begin{figure*}
    \centering
    \includegraphics[width=0.8\textwidth]{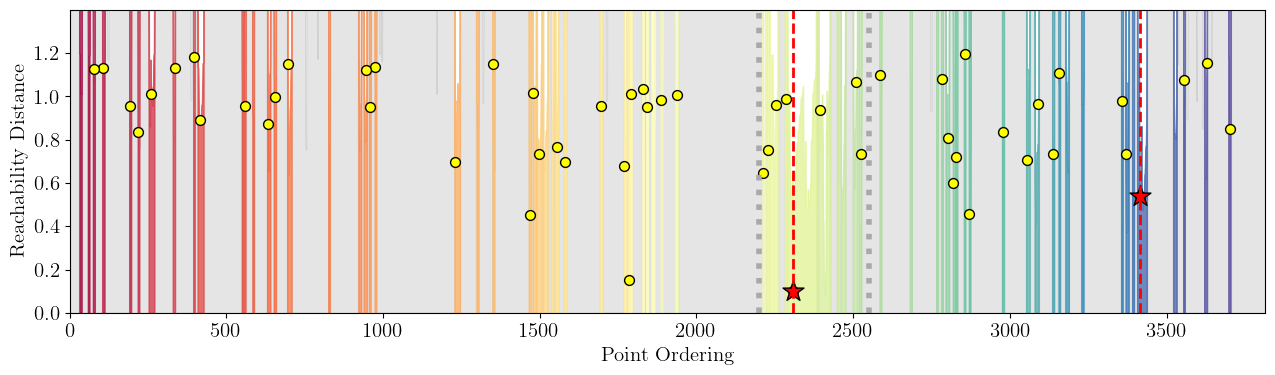}
    \includegraphics[width=0.8\textwidth]{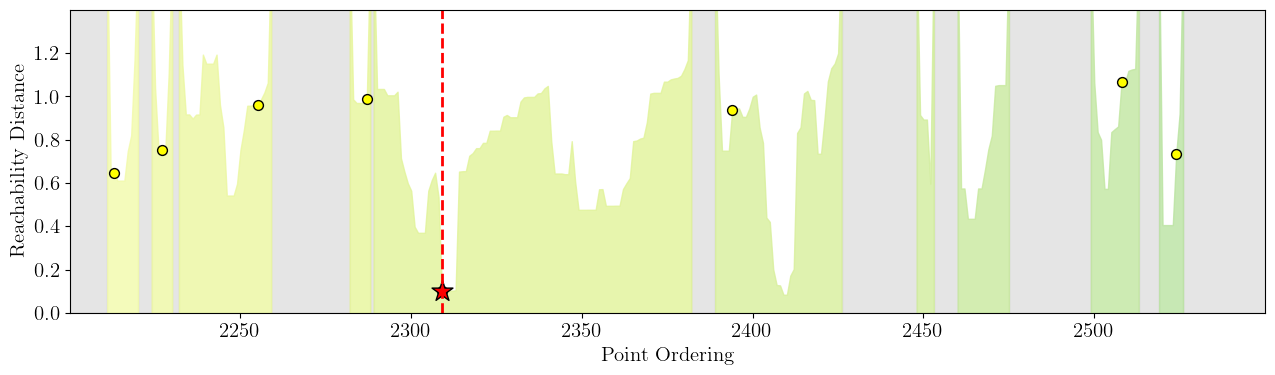}
    \caption{The reachability plot for galaxy groups from a sample covering a $30^2 \, \deg ^2$  sky area within the redshift range $0.7<z<0.8$. The colors highlight different clusters as determined by the clustering process described in Section \ref{sec:scaled-optics}. The top panel displays the complete reachability plot of this sample, while the bottom panel focuses on a specific section marked by grey dashed lines in the top panel. Light grey regions in each panel indicate isolated galaxies. Yellow points represent the brightest galaxies identified in each group. The red line and red star marker denote the reachability distance of the BCGs identified in \citetalias{hsuSDSSIVMaNGACannibalism2022} and those successfully recovered in our analysis, respectively.}
    \label{fig:RP-plot}
\end{figure*}

Using the best-fitted parameters and the LOS scaling factor listed in Table \ref{tab:hypervalue-optimal} and \ref{tab:hypervalue-s_los}, we successfully identified 115 BCGs consistent with the 118 BCGs identified in \citetalias{hsuSDSSIVMaNGACannibalism2022}. The spatial distribution of the galaxy clusters corresponding to these BCGs in redshift and right ascension (RA) space is illustrated in Figure \ref{fig:BCG-distribution}. Only three relatively small clusters failed to be predicted by sOPTICS. Figure \ref{fig:RP-plot} shows a segment of the reachability plot for sOPTICS, where the galaxy clusters appear as distinct, deep valleys. The gray areas represent isolated field galaxies that are significantly distanced from others.  The bottom panel shows a specific example cluster's reachability distances and neighbors, including a BCG recovered from the \citetalias{hsuSDSSIVMaNGACannibalism2022} sample. It is shown that these density-based clustering methods, such as OPTICS and sOPTICS, have given us a clear and straightforward picture of the position of BCGs in clusters. In this particular case, the BCG is located precisely at the densest part of the region, indicating a perfect alignment with the cluster's center of gravity. However, it is also evident from other commonly detected cases (highlighted as yellow spots in the plot) that the BCGs are not always situated at the densest part of the cluster. This variation highlights the diversity in the spatial distributions of BCGs and explains why our sOPTICS method did not successfully predict three BCGs.

\subsection{sOPTICS: a robust group and BCG finder}

One of the key strengths of sOPTICS is its ability to serve as a proxy for those complex physically motivated, FoF-based approaches while retaining efficiency and adaptability. By focusing on dense regions in a dataset, sOPTICS identifies galaxy clusters across varying environments with minimal tuning and computational overhead, as demonstrated by its high recovery rate relative to the \citetalias{yangGalaxyGroupsSDSS2007} catalog and other observational samples (Sections~\ref{sec:Shi}). These findings highlight sOPTICS’s utility in constructing reliable galaxy group catalogs for large redshift surveys. Even when parameters are not extensively recalibrated, sOPTICS can maintain strong performance, offering both practical scalability and solid clustering outcomes. This resilience across diverse datasets makes sOPTICS a promising, and efficient tool for future galaxy-group identification tasks, especially in scenarios where redshift-space distortions pose significant challenges. Unlike FoF, which requires carefully chosen linking lengths that can merge or dilute cluster boundaries, sOPTICS adaptively captures complex group structures and provides valuable information on internal density variations within clusters.

An additional advantage of sOPTICS lies in its natural compatibility with BCGs searches. Because BCGs generally reside near the densest region of a cluster, a clustering algorithm that prioritizes dense structures can be particularly effective at locating these galaxies. Our tests against the \citetalias{hsuSDSSIVMaNGACannibalism2022} sample confirm this potential (Section~\ref{sec:test-BCG}). For researchers specifically targeting BCGs, we recommend using sOPTICS with relatively small $\epsilon$, large $N_\mathrm{min}$, and reasonable $s_\mathrm{LOS}$ values to highlight very dense clusters. This setup streamlines the search for potential BCG candidates in expansive surveys, offering a time-efficient alternative to highly iterative or computationally intensive methods.

Nevertheless, while BCGs are often the most luminous galaxies within a cluster, they do not necessarily coincide with the cluster’s geometric center, given that spatial centering is influenced by halo geometry and luminosity distributions \citep{skibbaAreBrightestHalo2011}. Hence, accurately identifying a BCG requires a careful selection process that considers factors such as brightness, proximity to the cluster center, and corroborating information from multi-wavelength surveys. Verification with X-ray or optical follow-up can further constrain the reliability of a given BCG identification.

Taken together, our findings demonstrate that sOPTICS can be used effectively on both simulated and real datasets to identify large-scale structures and delve into their internal density patterns. Its efficiency, adaptability, and capacity for uncovering the densest regions within clusters make it a promising choice for both general group-finding purposes and specialized tasks such as BCG identification.

\section{Summary and Conclusion}
\label{sec:summary}

This study evaluated the effectiveness of eight popular clustering algorithms in data science for identifying galaxy groups and clusters through tests involving comparisons with both simulations and existing reliable group catalogs. 
Our findings indicate that our sOPTICS algorithm is a robust galaxy group finders. 
In particular, sOPTICS demonstrates significant flexibility in its hyperparameters and, when combined with a line-of-sight scaling factor to mitigate redshift-space distortion, exceeds FoF in both efficiency and accuracy. This advantage is especially apparent in pinpointing the densest regions of galaxy groups and identifying BCGs across large surveys.

We conclude that scaled OPTICS and FoF are comparably effective, with sOPTICS showing higher purity and recovery rates. While FoF can be faster and more computationally efficient, especially for large datasets -- a recognized asset for very large datasets from observation -- its performance heavily depends on the choice of linking length. Despite this dependency, FoF, as a popular and classical clustering method in astrophysics, remains particularly effective for low redshift surveys where redshift space distortion is less significant.

Our investigation highlights three principal strengths of sOPTICS:

\begin{itemize}
    \item Robustness to a wide range of hyperparameter values. We have identified two empirical relationships involving $\epsilon$, $N_\mathrm{min}$, and $s_\mathrm{LOS}$, which provide practical guidance for setting these parameters to achieve reliable clustering results.
    \item Unlike many clustering algorithms, sOPTICS does not immediately segment data into clusters but generates a reachability plot of distances to the nearest neighbors within the$ \epsilon$-neighborhood. Clusters are then identified as valleys in this plot. This design makes sOPTICS less sensitive to hyperparameter tuning, and, most notably, enables it to effectively capture the promising galaxy groups/clusters without any exhaustive parameter search.
    \item Focusing on the densest regions of the dataset, where BCGs frequently reside. By choosing extreme but physically motivated hyperparameter values, one can efficiently isolate the most significant clusters, making sOPTICS a practical method for surveying massive structures without requiring more complex group-finding computing.
\end{itemize}

Looking ahead, we anticipate leveraging richer and more precise galaxy data from observations such as the Dark Energy Spectroscopic Instrument \citep[DESI;][]{deyOverviewDESILegacy2019} and other large-scale spectroscopic surveys. These expanded datasets will facilitate more realistic hyperparameter modeling, especially for higher-redshift clusters, and enable further refinement of the empirical relationships suggested by $\epsilon$, $N_\mathrm{min}$, and$ s_\mathrm{LOS}$. By integrating these enhancements, sOPTICS may be developed into an even more powerful and adaptable tool for next-generation astrophysical surveys, allowing researchers to map cosmic structures quickly and reliably over vast regions of the observable universe.

\section*{Acknowledgements}

We thank the anonymous referee for the careful review and nice suggestions that have been incorporated into and improved the paper, as well as Yong Tian and Yi Duann for their helpful comments and discussions about BCGs.

This work has been supported by the Japan Society for the Promotion of Science (JSPS) Grants-in-Aid for Scientific Research (21H01128, 24H00247, and JP21J23611). 
This work has also been supported in part by the Collaboration Funding of the Institute of Statistical Mathematics ``New Perspective of the Cosmology Pioneered by the Fusion of Data Science and Physics''. 
\mbox{S.\, C.} has been supported by the JSPS Grant No. JP21J23611.

\section*{Data Availability}

The data utilized in this study is publicly accessible at the following sources: \url{http://sdss.physics.nyu.edu/vagc/} and \url{https://gax.sjtu.edu.cn/data/Group.html}. Additionally, the code and derived data produced during this research can be made available upon reasonable request to the first author.



\bibliographystyle{mnras}
\bibliography{Clustering} 






\bsp	
\label{lastpage}
\end{document}